\documentclass[12pt]{article}
\usepackage{cite}
\usepackage{enumerate}
\usepackage{ifpdf}
\usepackage{ae}
\usepackage[T1]{fontenc}
\usepackage[ansinew]{inputenc}
\usepackage{amsmath}
\usepackage{relsize}
\usepackage{amssymb}
\usepackage{tikz}
\usepackage{graphicx}
\usepackage{color}
\definecolor{darkblue}{cmyk}{0.9,0.9,0,0}
\definecolor{darkgreen}{rgb}{0,0.55,0}
\usepackage[colorlinks=true,linkcolor=darkblue,citecolor=darkblue,urlcolor=darkblue]{hyperref}
\usepackage{epsfig}
\usepackage{epstopdf}
\usepackage{graphicx}

\makeatletter
\long\def\@makecaption#1#2{
  \vskip\abovecaptionskip
  \sbox\@tempboxa{{\captionfonts #1: #2}}
  \ifdim \wd\@tempboxa >\hsize
    {\captionfonts #1: #2\par}
  \else
    \hbox to\hsize{\hfil\box\@tempboxa\hfil}
  \fi
  \vskip\belowcaptionskip}
\makeatother

\def\twp{{\cal O}_{20'}}
\def\coll{{\rm coll}}
\def\eps{\epsilon}
\def\reg{{\rm reg}}

\def\tree{{\rm tree}}
\def\mtree{M_{\rm tree}}
\def\mloop{M_{\rm 1-loop}}
\def\Gc{{\cal G}}
\def\z{\zeta}
\def\d{\delta}

\def\0{{(0)}}
\def\1{{(1)}}
\def\2{{(2)}}
\def\3{{(3)}}
\def\4{{(4)}}

\newcommand{\Res}[1] {\underset{#1}{\text{Res}}\,}

\def\Q{{\cal Q}}

\newcommand{\e}[2] {\begin{equation} \label{#1} #2 \end{equation}}
\newcommand{\es}[2] {\begin{equation} \label{#1} \begin{split} #2 \end{split} \end{equation}}
\def\eqr{\eqref}
\def\sec{\section}
\def\subsec{\subsection}

\def\l{\lambda}
\def\vs{\vskip .1 in}

\def\a{\alpha}
\def\rar{\rightarrow}

\def\la{\langle}
\def\ra{\rangle}
\def\O{{\cal O}}
\def\OO{[\O\O]}
\def\i{\infty}
\def\p{\partial}

\def\ssec{\subsection}
\def\sssec{\subsubsection}
\def\sec{\section}

\def\vs{\vskip .1 in}
\def\D{\Delta}

\newcommand{\beq}{\begin{equation}}
\newcommand{\eeq}{\end{equation}}
\newcommand{\beqy} {\begin{eqnarray}}
\newcommand{\eeqy} {\end{eqnarray}}
\newcommand{\bsmat}{\begin{smallmatrix}}
\newcommand{\esmat}{\end{smallmatrix}}
\newcommand{\bmat}{\begin{matrix}}
\newcommand{\emat}{\end{matrix}}

\def\({\left(}
\def\){\right)}
\def\[{\left[}
\def\]{\right]}

\def\<{\langle}
\def\>{\rangle}

\def\a{\alpha}

\def\g{\gamma}
\def\G{\Gamma}
\def\d{\delta}
\def\z{\zeta}

\def\l{\lambda}

\def\t{\tau}

\def\cO{{\cal O}}

\usepackage{varioref}
\usepackage{makeidx}
\makeindex

\usepackage[english]{babel}
\usepackage{caption}
\usepackage{subfig}

\usepackage{tabularx}
\begin{document}

\thispagestyle{empty}

\renewcommand{\thefootnote}{\fnsymbol{footnote}}
\setcounter{page}{1}
\setcounter{footnote}{0}
\setcounter{figure}{0}

\begin{titlepage}

\begin{center}

\vskip 2.3 cm 

\vskip 5mm

{\Large \bf 
Loops in AdS from Conformal Field Theory}
\vskip 0.5cm

\vskip 15mm

\centerline{Ofer Aharony$^{a}$, Luis F. Alday$^{b}$, Agnese Bissi$^{c}$ and Eric Perlmutter$^d$}
\bigskip
\centerline{\it $^{a}$ Department of Particle Physics and Astrophysics, Weizmann Institute of Science,}
\centerline{\it Rehovot 7610001, Israel}
\vs
\centerline{\it $^{b}$ Mathematical Institute, University of Oxford,} 
\centerline{\it  Andrew Wiles Building, Radcliffe Observatory Quarter,}
\centerline{\it Woodstock Road, Oxford, OX2 6GG, UK}
\vs
\centerline{\it $^{c}$ Center for the Fundamental Laws of Nature,}
\centerline{\it Harvard University, Cambridge, MA 02138 USA}
\vs
\centerline{\it $^{d}$ Department of Physics, Princeton University}
\centerline{\it  Jadwin Hall, Princeton, NJ 08544 USA}

\end{center}

\vskip 2 cm

\begin{abstract}
\noindent We propose and demonstrate a new use for conformal field theory (CFT) crossing equations in the context of AdS/CFT: the computation of loop amplitudes in AdS, dual to non-planar correlators in holographic CFTs. Loops in AdS are largely unexplored, mostly due to technical difficulties in direct calculations. We revisit this problem, and the dual $1/N$ expansion of CFTs, in two independent ways. The first is to show how to explicitly solve the crossing equations to the first subleading order in $1/N^2$, given a leading order solution. This is done as a systematic expansion in inverse powers of the spin, to all orders. These expansions can be resummed, leading to the CFT data for finite values of the spin. Our second approach involves Mellin space. We show how the polar part of the four-point, loop-level Mellin amplitudes can be fully reconstructed from the leading-order data. The anomalous dimensions computed with both methods agree. In the case of $\phi^4$ theory in AdS, our crossing solution reproduces a previous computation of the one-loop bubble diagram. We can go further, deriving part of the four-point function in $\phi^3+\phi^4$ theory in AdS which had never been computed. In the process, we show how to analytically derive anomalous dimensions from Mellin amplitudes with an infinite series of poles, and discuss applications to more complicated cases such as the ${\cal N}=4$ super-Yang-Mills theory. 

\end{abstract}

\end{titlepage}

\setcounter{page}{1}
\renewcommand{\thefootnote}{\arabic{footnote}}
\setcounter{footnote}{0}

\newpage
\setcounter{tocdepth}{2}
\tableofcontents

 \def\nref#1{{(\ref{#1})}}

\section{Introduction}

The AdS/CFT Correspondence is, in its most well-studied form, a duality between weakly coupled theories of gravity in anti-de Sitter (AdS) space and conformal field theories (CFTs) with many degrees of freedom (``large $N$''). Perhaps the most fundamental element in the holographic dictionary is that the AdS path integral with boundary sources is the generating function of dual CFT correlation functions, thus making predictions for large $N$, typically strongly coupled, dynamics. The $1/N$ expansion of the CFT correlators maps to the perturbative expansion of AdS amplitudes, which is computed via the loop expansion of Witten diagrams \cite{Maldacena:1997re,Gubser:1998bc,Witten:1998qj}.

Such basics may seem hardly worth stressing: conceptually, the AdS side of this story appears rather straightforward, and no different from flat space. However, perhaps surprisingly, the physical content of the AdS perturbative expansion is poorly understood. Beyond tree-level, the computation of AdS amplitudes is nearly unexplored, as almost nothing has been computed. At one-loop and beyond, technical challenges inhibit brute force position space computations: simple one-loop diagrams whose flat space counterparts appear in introductory quantum field theory courses, like the three-point scalar vertex correction and the four-point scalar box diagram, have not been computed in AdS. Even at tree-level, the original calculations \cite{Muck:1998rr,Liu:1998ty, Freedman:1998bj,  Liu:1998th, D'Hoker:1998mz, D'Hoker:1999pj, D'Hoker:1999jp, D'Hoker:1999ni, Hoffmann:2000mx, Arutyunov:2002fh} were impressive but arduous, and struggled to make manifest the relation to CFT data; only recently have leaner, more transparent methods been introduced, including Mellin space \cite{0907.2407, Penedones:2010ue} and geodesic Witten diagrams \cite{Hijano:2015zsa}. We emphasize that these are not related to challenges of coupling to gravity: in an AdS effective field theory sans gravity, that can be dual to a decoupled sector of some CFT \cite{Aharony:2015zea}, the same issues are present.  

The purpose of this paper is to initiate a systematic exploration of loop amplitudes in AdS, and of the dual $1/N$ expansion of holographic CFT correlation functions, using modern methods. 
 \begin{figure}[t!]
   \begin{center}
 \includegraphics[width = \textwidth]{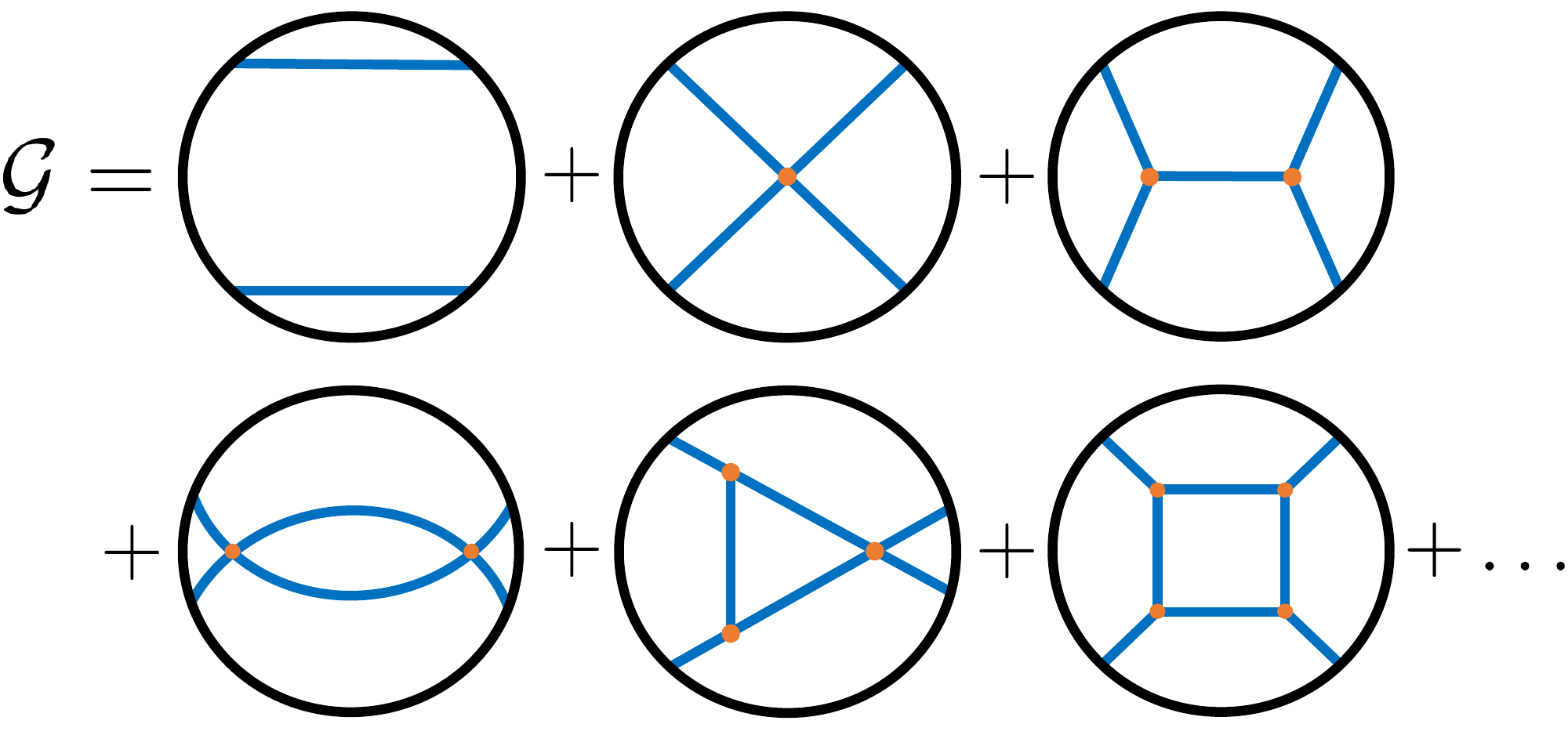}
 \caption{The schematic form of the loop expansion of AdS four-point amplitudes ${\cal G}$, shown only in a single channel for simplicity. The one-loop diagrams are holographically dual to the leading non-planar corrections to four-point correlation functions in holographic, large $N$ CFTs.}
 \label{loopexp}
 \end{center}
 \end{figure}

There are (at least) two main reasons why one might be interested in this problem. The first is to understand the structure of amplitudes in curved space, and in AdS in particular. For inspiration and contrast, consider the decades of fantastic progress in understanding flat space S-matrices, which contain extremely powerful physical and mathematical structures: they relate loops to trees \cite{Bern:1994zx,Bern:1994cg}, gravitational theories to gauge theories \cite{Bern:2008qj}, and have suggested a re-imagination of the role played by spacetime itself \cite{Arkani-Hamed:2013jha}. One is led to ask: What is the organizing principle underlying the structure of AdS scattering amplitudes? Given the existence of a well-defined flat space limit of AdS (Mellin) amplitudes \cite{Penedones:2010ue, Paulos:2016fap}, the aforementioned structures should be encoded in, or extend to, the analogous AdS amplitudes. 

The second is to better understand the large $N$ dynamics of holographic CFTs. The marvelous universality of holographic large $N$ CFTs is typically only studied at leading order, dual to classical calculations in AdS. But the definition of a holographic CFT must hold at every order in the $1/N$ expansion. For instance, a large $N$ CFT whose entanglement entropy obeys the Ryu-Takayanagi formula \cite{Ryu:2006bv}, but not the Faulkner-Lewkowycz-Maldacena correction term \cite{Faulkner:2013ana}, cannot be dual to Einstein gravity coupled to matter. It is the analogous correction that we would like to understand about the CFT operator product expansion (OPE) data: namely, what the loop-level constraints are on operator dimensions and OPE coefficients due to the existence of a weakly coupled gravity dual. In addition, for given holographic CFTs whose planar correlation functions are known, we would like to understand how to go to higher orders in the $1/N$ expansion.

While there is some work on one-loop AdS amplitudes \cite{Cornalba:2007zb,Penedones:2010ue,Fitzpatrick:2011hu, Fitzpatrick:2011dm}, some of which we will make contact with later, loop physics in AdS has mostly been studied using other simpler observables, specifically the partition function (e.g. \cite{Mansfield:2000zw, Giombi:2008vd, David:2009xg, Giombi:2013fka, Ardehali:2013xya,Beccaria:2014xda}). Interesting constraints can indeed be extracted from the one-loop partition function -- for example, in a four-dimensional CFT, $1/N$ corrections to $a$ and $c$ can be computed by adding Kaluza-Klein contributions to the Casimir energy in global AdS$_5$ -- but correlation functions are much richer objects. In particular, they depend on OPE coefficients and coordinates, and can hence access Lorentzian regimes of CFT. Knowing loop amplitudes in a given bulk theory would open the door to non-planar extensions of dynamical aspects of holography and the conformal bootstrap \cite{ElShowk:2011ag, Fitzpatrick:2012yx, Komargodski:2012ek,  Fitzpatrick:2014vua, Camanho:2014apa, Alday:2014tsa, Maldacena:2015waa, Maldacena:2015iua, Hartman:2015lfa, Li:2015rfa, Hartman:2016dxc,Komargodski:2016gci,Perlmutter:2016pkf, Hofman:2016awc, Alday:2016htq, Hartman:2016lgu, Afkhami-Jeddi:2016ntf}.

While AdS loop amplitudes apparently pose difficult technical problems in position space, there is reason for optimism. From the AdS point of view, given a classical effective action corresponding to the leading order in the $1/N$ expansion, one extracts the Feynman rules, and computes loop diagrams accordingly. In this sense, loop amplitudes are fixed upon knowing all tree amplitudes, in principle. More precisely, the results of loop computations are uniquely determined up to the need to fix renormalization conditions for some parameters; for any theory, renormalizable or not, only a finite number of conditions is required at any given loop order. The problem is to make the relation between loop-level and tree-level AdS amplitudes precise, \`a la the Feynman tree theorem and generalized unitarity methods for S-matrices. 

How can quantitative progress be made? We will show that analytic solutions of the conformal bootstrap for these four-point functions may be found at subleading orders in the $1/N$ expansion. This may be viewed as either a CFT or a bulk calculation. The leading order solutions for the connected four-point function of a single $\mathbb{Z}_2$-invariant scalar primary were constructed in \cite{Heemskerk:2009pn}, where they showed that there is a one-to-one mapping between those solutions and classical scalar field theories on AdS space with local quartic interactions. An important technical simplification of the leading order solutions is that they have finite support in the spin, which makes manifest the analytic properties of the four-point correlator. At subleading order this is no longer the case and the method of \cite{Heemskerk:2009pn} does not apply. Nevertheless, solutions can be constructed  as a systematic expansion around large spin, adapting the machinery of \cite{Alday:2015eya, Alday:2015ewa, Alday:2016njk}. We find that the solution to order $1/N^4$ is fully fixed in terms of the data to order $1/N^2$, to all orders in the inverse spin expansion. Likewise we will show that the Mellin representation of the CFT four-point functions makes it clear why and how higher orders in $1/N$ are determined by the leading-order result. Furthermore, we will reconstruct the full one-loop Mellin amplitude for several examples.

\ssec{Setup}

Throughout the paper, we study an identical-scalar four-point function, $\la \O\O\O\O\ra$, for a scalar primary $\O$ of dimension $\Delta$. This is determined in terms of an ``amplitude'' ${\cal G}(u,v)$ of the two conformal cross-ratios, $u$ and $v$ (more details will be given in the next section). $\Gc(u,v)$ admits an expansion in $1/N$:
 \begin{equation}
 {\cal G}(u,v) =  {\cal G}^{(0)}(u,v)+ \frac{1}{N^2}  {\cal G}^{(1)}(u,v)+ \frac{1}{N^4}  {\cal G}^{(2)}(u,v) + \cdots
 \end{equation}
The disconnected piece $\Gc^{\0}$ is determined by mean field theory, while $\Gc^{\1}\equiv \Gc_{\rm tree}$ and $\Gc^{\2}\equiv \Gc_{\rm 1-loop}$ are the bulk tree-level and one-loop amplitudes, respectively.\footnote{Following \cite{Heemskerk:2009pn}, we use the large $N$ gauge theory notation $1/N^2$ to stand for the small coupling in the bulk. In a generic, full-fledged holographic CFT, this stands for $1/c$ (even when $c$ does not scale as $N^2$). In a general bulk theory such as $\l\phi^4$, this stands for the four-point couplings (such as $\l$), while three-point bulk couplings scale as $1/N$. We will freely interchange the labels tree-level/first order/$O(1/N^2)$ throughout the paper, and likewise for one-loop/second order/$O(1/N^4)$.} At every order in the $1/N$ expansion, the amplitude is subject to the crossing equation
\e{}{v^{\D}\Gc^{(i)}(u,v) = u^{\D}\Gc^{(i)}(v,u)~.}
We will also work with the Mellin representation of this amplitude, $M(s,t)$, which admits an analogous expansion.  

Any large $N$ CFT containing $\O$ also necessarily contains a tower of ``multi-trace'' primary operators that are composites of $\O$. The most familiar of these are the double-trace operators $\OO_{n,\ell}$, one for each pair $(n,\ell)$ \cite{Penedones:2010ue}, whose definition we recall below. These acquire corrections to their conformal dimensions $\D_{n,\ell}$ and squared OPE coefficients $a_{n,\ell} \equiv C_{\O\O\OO_{n,\ell}}^2$, at every order in the $1/N$ expansion:
\es{}{ \Delta_{n,\ell} &= 2\Delta+2n+\ell+ \frac{1}{N^2} \gamma_{n,\ell}^{(1)}+  \frac{1}{N^4} \gamma_{n,\ell}^{(2)} + \cdots,\\
 a_{n,\ell} &= a_{n,\ell}^{(0)}+ \frac{1}{N^2} a_{n,\ell}^{(1)} + \frac{1}{N^4} a_{n,\ell}^{(2)}+ \cdots}
 Mean field theory determines $a_{n,\ell}^{(0)}$, and the tree-level crossing equation determines $\g^{\1}_{n,\ell}$ and $a^{\1}_{n,\ell}$ \cite{Heemskerk:2009pn}. To solve the one-loop crossing equation is to derive $\g^{\2}_{n,\ell}$ and $a^{\2}_{n,\ell}$, in addition to the OPE data for any other operators appearing at that order. We call the $\OO$ contributions to the one-loop Mellin amplitude $\mloop^{\OO}$.

\ssec{Summary of results}
In \cite{Heemskerk:2009pn}, the authors considered generalized free field sectors of holographic CFTs in which the only operators appearing at $O(1/N^2)$ are the $\OO$ double-trace operators. Such setups are dual to the simplest effective field theories in AdS, namely, $\phi^4$-type theories with no cubic couplings. We will sometimes call these theories ``truncated'' theories on account of the spin truncation $\g^{\1}_{n,\ell>L}=0$ for some finite $L$, as used in \cite{Heemskerk:2009pn}; a bulk theory with $2p$ derivatives at the vertices has $L=2\lfloor {p\over 2}\rfloor$. We note that in a truncated theory, the double-traces are the {\it only} contributions to the full $\mloop$: even at $O(1/N^4)$, the $\O\times\O$ OPE contains no single-trace operators by design, and no higher multi-trace operators by necessity (a fact which we explain in Appendix \ref{crossing_ops}).  

One advantage of Mellin space is that it allows us to show explicitly how $\mloop^{\OO}$ can in principle be derived directly from $1/N$ considerations alone. We show how large $N$ fixes the poles and residues of $\mloop^{\OO}$, for any theory, in terms of the tree-level anomalous dimensions $\g^{\1}_{n,\ell}$. The location of the poles has been understood, and the residues derived in a specific example, in \cite{Penedones:2010ue,Fitzpatrick:2011hu, Fitzpatrick:2011dm}; we show how to obtain this in general. We derive the leading residue explicitly for a general theory (see \eqr{R0s}). In a truncated theory, the leading residue is sufficient to determine the large spin asymptotics of $\g^{\2}_{n,\ell}$. The latter passes a check against the lightcone bootstrap \cite{Komargodski:2012ek, Fitzpatrick:2012yx} as applied to $\phi^4$ theory. 

However, the above approach is somewhat clunky to implement and is not maximally physically transparent. A more elegant, and more practical, approach is to solve the one-loop crossing equations for 
 $\g^{\2}_{n,\ell}$ and $a^{\2}_{n,\ell}$. This is tantamount to knowing the dual AdS one-loop amplitude. The statement of bulk reconstruction is not just philosophical: we can actually reconstruct $\mloop$ from OPE data, because they are related by a linear Mellin integral transform. {\it This is our proposed use for crossing symmetry: given leading order OPE data, we solve the crossing equations at the next order, thus reconstructing $\mloop$ for the dual AdS theory.}

Let us now discuss what is involved in actually solving the loop-level crossing equations. At one-loop, the tree-level data acts as a source in the crossing equation for $\Gc^{\2}(u,v)$, which has a unique inhomogeneous solution. The freedom to add a homogeneous solution matches expectations from the bulk, where one is free to modify the local quartic couplings at every loop order: from \cite{Heemskerk:2009pn}, the correspondence between local quartic vertices and homogeneous solutions to crossing follows. This pattern continues at higher orders.

In this work, we will focus on the anomalous dimensions $\g^{\2}_{n,\ell}$. To actually compute these from crossing, our main observation may be sketched as follows. In the regime $u\ll v\ll1$, $\Gc^{\2}(u,v)$ contains terms of the form
\e{}{\Gc^{\2}(u,v) \supset u^\Delta f(u)\log^2 (u) \log v~,}
where $f(u)$ is fixed by lower-order data, and is quadratic in the first-order anomalous dimensions $\g^{\1}_{n,\ell}$ (hence the $\log^2 (u)$). By crossing symmetry, we also have 
\e{g2cros}{\Gc^{\2}(u,v) \supset u^\Delta f(v)\log^2 (v) \log u~.}
At this stage we specify to truncated theories, where $\lbrace a^{\1}_{n,\ell},\g^{\1}_{n,\ell}\rbrace$ vanish above some finite $\ell$. It is easy to see from the small $u$ expansion that the term in \eqr{g2cros} must come from a contribution to $\Gc^{\2}(u,v)$ that is linear in $\g^{\2}_{n,\ell}$. For $n=0$, where the analysis is simplest, the precise equation is
\e{ceqn}{\sum_\ell a_{0,\ell}^{(0)} \gamma_{0,\ell}^{(2)} g_{2\Delta+\ell,\ell}^\coll(v) = 2 f(v) \log^2 (v)+ \cdots}
where $g_{2\Delta+\ell,\ell}^\coll(v)$ is the lightcone, or collinear, conformal block, and ``$\cdots$'' denotes logarithmically divergent or regular terms. This is the desired equation for $\g^{\2}_{0,\ell}$ in terms of first-order data $\g^{\1}_{0,\ell}$.

The solution of \eqr{ceqn} is performed order-by-order in the large spin expansion: because each term on the left-hand side diverges like $\log v$, it must be that $\g^{\2}_{0,\ell}\neq 0$ for all $\ell$, and its large spin behavior is determined by matching to $f(v)$. At leading order, one finds $\g^{\2}_{0,\ell} \sim \ell^{-2\D}$. A systematic expansion requires further development of the Casimir methods utilized in \cite{Alday:2015eya, Alday:2015ewa, Alday:2016njk}, adapted now to this particular one-loop equation. Given a large spin expansion of $\g^{\2}_{0,\ell}$, a resummation down to {\it finite} spin is possible when $\D\in\mathbb{Z}$. Altogether, both the large and finite spin data constitute a holographic construction of the one-loop amplitudes in the dual AdS theory that classically gives rise to the $\g^{\1}_{0,\ell}$ used in the crossing problem. 

In the above large spin analysis, we encounter an exciting mathematical surprise: a certain class of harmonic polylogarithms forms a basis of solutions. In particular, if we expand $\g^{\2}_{0,\ell}$ to $n^{\rm th}$ order in inverse powers of the collinear Casimir eigenvalue $J^2 = (\ell+\D)(\ell+\D-1)$, then for integer $\D>1$, $f(v)$ can be written as a linear combination of weight $w\leq \D-2+n$ harmonic polylogs, defined in \eqr{hpl1}-\eqr{hpl2}.\footnote{The basis for $\D\notin\mathbb{Z}$ can be thought of as comprised of analytic continuations of harmonic polylogs to non-integer weight. It would be interesting to formalize this.} Harmonic polylogs are specified by a weight vector, and only a specific subclass of such functions appears in our problem, namely those specified by the alternating $w$-vector $\vec\rho_w = (\ldots,0,1,0,1)$. Given that multiple polylogs are ubiquitous in one-loop amplitudes in flat space, it is intriguing to see some of them appearing in the construction of one-loop amplitudes in AdS via the crossing equations.

Before showing our results for specific theories, we should address an obvious question: what happens when an AdS theory has a UV divergence? In particular, how is this visible in the solutions to crossing? This has a satisfying answer. We expect to be able to cancel UV divergences by adding a finite number of local counterterms to our AdS effective action at a given loop order, just as in flat space. As explained in \cite{Heemskerk:2009pn}, local quartic vertices with $2p$ derivatives generate anomalous dimensions only for double-trace operators of spin $\ell\leq2\lfloor {p\over 2}\rfloor$. Therefore, on account of bulk locality of the divergences, we have a precise prediction: when we compute a divergent one-loop bulk diagram via crossing, $\g^{\2}_{n,\ell}$ should diverge for the above range of spins, where $p$ is the number of derivatives in the counterterm. Moreover, for any regularization, the divergence should be proportional to $\g^{\1}_{n,\ell}$. Analogous statements apply at any loop order.

We demonstrate all of the above explicitly in the following two examples:

{\bf 1) $\phi^4$ in AdS.} The only non-trivial one-loop diagram is the bubble diagram of Figure \ref{loopexp}. This is the one case where $\mloop$ is actually known directly from a bulk calculation, performed in Mellin space in \cite{Fitzpatrick:2011hu}: the authors used an AdS analog of the K\"all\'en-Lehmann representation to write the loop as an infinite sum of trees. Using our large spin data, we reconstruct this amplitude in AdS$_3$ and AdS$_5$ for a $\D=2$ scalar, exactly matching the result of \cite{Fitzpatrick:2011hu}. (See \eqr{mloopform}-\eqr{d2res}.) 

Moreover, we show how to analytically compute $\g^{\2}_{0,\ell}$ at some {\it finite} spins directly from $\mloop$ itself. The results match the resummation of the large spin solutions to crossing. This extraction had not been done previously -- indeed, we know of no case in the literature where OPE data has been analytically derived from a Mellin amplitude with an infinite series of poles. We expect our regularization techniques to be useful more widely in the world of Mellin amplitudes. The results at low spin align precisely with our UV divergence expectations. The AdS$_3$ theory is finite, but the AdS$_5$ theory requires a $\phi^4$ counterterm. Accordingly, $\g^{\2}_{0,0}$ diverges in $d=4$ (AdS$_5$) but not in $d=2$ (AdS$_3$), and $\g^{\2}_{0,\ell}$ for $\ell=2,4$ is finite in both cases. (See \eqr{d2l0}-\eqr{d4l4}.)
 
{\bf 2) ${\mu_3\over 3!}\phi^3+{\mu_4\over 4!}\phi^4$ in AdS.} We compute the part of the four-point amplitude which involves vertices of each type, proportional to $\mu_3^2\mu_4$. This includes the triangle, shown in Figure \ref{loopexp}. This has never been computed, in any bulk spacetime dimension, as the trick of \cite{Fitzpatrick:2011hu} does not work here. Taking $\D=2$ for concreteness, we compute large and finite spin anomalous dimensions from crossing (see \eqr{phi3ad}-\eqr{largeJp34}), and reconstruct $\mloop$. In $d=4$, and in the $t$-channel, say,
\e{mlooptri}{\mloop(s,t) = \left(\frac{40 \, _3F_2\left(1,1,2-\frac{t}{2};\frac{5}{2},3-\frac{t}{2};1\right)}{t-4}+\frac{56 \, _3F_2\left(2,2,3-\frac{t}{2};\frac{7}{2},4-\frac{t}{2};1\right)}{5 (t-6)}\right)\mu_3^2\mu_4~.}
After subtracting diagrams proportional to $\mu_3^2\mu_4$ that renormalize the tree-level exchange, which may be computed using existing technology, this yields the holographic computation of the four-point triangle Witten diagram in AdS$_5$ for a $m^2=-4$ scalar. We emphasize that our approach gives the complete, crossing-symmetric amplitude at order $\mu_3^2\mu_4$, without splitting it into Witten diagrams. We also give the analogous result in AdS$_3$ in \eqref{mlooptrid2}. It would be very interesting to discover new tools for a direct evaluation in the bulk. 

Overall, our work takes a step toward the finite $N$, Planckian regime by illuminating the structure of the perturbative amplitude expansion in AdS and in large $N$ CFT. As we discuss in Section \ref{s7}, we believe that there is potential for the large $N$ bootstrap to address interesting questions beyond the realm of holography.  

The paper is organized as follows. In Section \ref{crossingloops}, we set up the crossing problem and identify the key one-loop constraint. In Section \ref{s3}, we review the basics of Mellin amplitudes, and use large $N$ alone to explain how $\mloop$ is constrained by tree-level data, and to construct the leading residue explicitly. In Section \ref{s4}, we develop the necessary tools for solving the one-loop crossing equations in general. In Section \ref{s5}, we apply our machinery to compute the bubble diagram of $\phi^4$ in AdS, and the triangle diagram of $\phi^3+\phi^4$ in AdS, via crossing. In Section \ref{s6}, we explain quite generally how to compute low-spin anomalous dimensions from Mellin amplitudes with an infinite series of poles; as an example, we apply this to the one-loop bubble diagram in $\phi^4$. We conclude in Section \ref{s7} with a discussion of generalizations, applications to full-fledged CFTs like ${\cal N}=4$ super-Yang-Mills and the $d=6$ (2,0) theory, and other future directions. Some appendices include further details.

\section{Crossing symmetry in the $1/N$ expansion}
\label{crossingloops}

\subsection{Setup}
Consider a generic CFT with a large $N$ expansion and a large mass gap. More precisely, we assume there exists a ``single-trace" scalar operator ${\cal O}$ of dimension $\Delta$, and that all other single trace operators acquire a very large dimension as $N$ becomes large.  This is equivalent to considering a weakly coupled theory of a single scalar field in AdS, with three-point couplings proportional to $1/N$ and four-point couplings proportional to $1/N^2$.

Consider the four-point function of identical operators ${\cal O}$. Conformal symmetry implies 
\begin{equation}\label{4pf}
\langle {\cal O}(x_1) {\cal O}(x_2) {\cal O}(x_3) {\cal O}(x_4) \rangle = \frac{{\cal G}(u,v)}{x_{12}^{2 \Delta} x_{34}^{2\Delta}}   \,,
 \end{equation}
where $x_{ij}\equiv x_i-x_j$ and we have introduced the cross-ratios $u\equiv \frac{x_{12}^2x_{34}^2}{x_{13}^2x_{24}^2}$ and $v\equiv \frac{x_{14}^2x_{23}^2}{x_{13}^2x_{24}^2}$. Crossing symmetry implies
\begin{equation}
\label{crossing}
v^\Delta {\cal G}(u,v) = u^\Delta {\cal G}(v,u).
 \end{equation}
We would like to study solutions to the crossing equation in a large $N$ expansion, up to $O(1/N^4)$. As discussed in Appendix \ref{crossing_ops}, up to this order and to inverse powers of the mass gap the operators appearing in the OPE of ${\cal O}$ with itself in a generic CFT are
\begin{equation}
{\cal O} \times {\cal O} = 1+ {\cal O}+ T_{\mu \nu} + [{\cal O}{\cal O}]_{n,\ell} + [TT]_{n,\ell} + [\cO T]_{n,\ell}~,
\end{equation}
where $1$ denotes the identity operator, $T$ the stress tensor, and the double-trace operators $[{\cal O}{\cal O}]_{n,\ell}$ are conformal primaries of the schematic form $[{\cal O}{\cal O}]_{n,\ell} = {\cal O} \Box^n \partial_{\mu_1} \cdots \partial_{\mu_\ell} {\cal O}+\cdots$. The presence of ${\cal O}$ and $[\O T]_{n,\ell}$ is forbidden in a theory with $\mathbb{Z}_2$ symmetry. Furthermore, in the simplest setting we can ignore the presence of the operators including the stress tensor; this is a good approximation when the self-couplings of the scalar are much larger than its gravitational couplings, which is true in particular for a non-gravitational theory on AdS. On the other hand, the presence of double-trace operators $[{\cal O}{\cal O}]_{n,\ell}$ is necessary for consistency with crossing symmetry. Note that higher-trace operators will appear at higher orders in the $1/N$ expansion, but not at $O(1/N^4)$. 

Let us for the moment focus on the simplest setting, in which the operators in the OPE include only the identity operator and double trace operators $[{\cal O}{\cal O}]_{n,\ell}$. This is relevant for computing correlators of a $\phi^4$ theory in AdS. In this case the four-point function admits the following conformal partial waves decomposition:
\begin{equation}
\label{CPW}
{\cal G}(u,v) =1+ \sum_{n=0}^\i\sum_{\ell~\text{even}}^\i a_{n,\ell} u^{\frac{\Delta_{n,\ell}-\ell}{2}} g_{\Delta_{n,\ell},\ell}(u,v) \,,
 \end{equation}
in which only even values of $\ell$ appear,\footnote{We assume that the identical external operators are uncharged under any global symmetries. Henceforth we leave the even spin restriction implicit, and use $\sum\limits_{n,\ell} \equiv \sum\limits_{n=0}^\i\sum\limits_{\ell\,\text{even}}^\i$.} and $a_{n,\ell}$ denote the OPE coefficients squared of $[{\cal O}{\cal O}]_{n,\ell}$ in the ${\cal O} \times {\cal O}$ OPE. The normalization of ${\cal O}$ has been chosen such that the contribution of the identity operator is exactly $1$. The conformal block for exchange of a dimension $\D_p$, spin-$\ell$ primary is written as
\e{Gblock}{G_{\D_{p},\ell}(u,v) = u^{\D_{p}-\ell\over 2}g_{\D_{p},\ell}(u,v)}
so as to make manifest the leading behaviour for small $u$. Although most of the methods of this paper will be general, we will mostly focus on $d=2$ and $d=4$ for definiteness. For these cases the conformal blocks are given by \eqr{Gblock} with
 \begin{align}
 g_{\D_p,\ell}(z,\bar z) &= \frac{z^\ell F_{\frac{\D_p+\ell}{2}}( z)F_{\frac{\D_p-\ell}{2}}(\bar z)+\bar z^\ell F_{\frac{\D_p-\ell}{2}}( z)F_{\frac{\D_p+\ell}{2}}(\bar z)}{1+\delta_{\ell,0}},~~~~~~~~~~~~~\text{for $d=2$} \\
  g_{\D_p,\ell}(z,\bar z) &= \frac{z^{\ell+1} F_{\frac{\D_p+\ell}{2}}( z)F_{\frac{\D_p-\ell-2}{2}}(\bar z)-\bar z^{\ell+1} F_{\frac{\D_p-\ell-2}{2}}( z)F_{\frac{\D_p+\ell}{2}}(\bar z)}{z-\bar z},~~~\text{for $d=4$} 
 \end{align}
 where we have introduced the parametrization $u=z \bar z, v= (1-z)(1-\bar z)$ for the cross-ratios, and $F_{\beta}(z)\equiv~_2F_1(\beta,\beta,2\beta;z)$. 
 
 At zeroth order in a $1/N$ expansion the four-point correlator \eqref{4pf} is simply the sum over the disconnected contribution in all three channels:
 \begin{equation}
 {\cal G}^{(0)}(u,v) = 1 + u^\Delta+ \left(\frac{u}{v}\right)^\Delta\,.
 \end{equation}
 This is consistent with the expected spectrum for double-trace operators at zeroth order
\begin{equation}
\Delta_{n,\ell}^{(0)}=2\Delta+2n+\ell \,,
\end{equation}
and leads to the following OPE coefficients \cite{Heemskerk:2009pn}
\begin{align}
a_{n,\ell}^{(0)} &= 2 C_n^{\Delta} C_{n+\ell}^{\Delta},~~~&\text{for $d=2$}\\
a_{n,\ell}^{(0)} &= \frac{2(\ell+1)(2\Delta+2n+\ell-2)}{(\Delta-1)^2} C_n^{\Delta-1} C_{n+\ell+1}^{\Delta-1},~~~&\text{for $d=4$}
\end{align}
where we have introduced 
\begin{equation}
 C_n^{\Delta} = \frac{\Gamma^2(\Delta+n)\Gamma(2\Delta+n-1)}{\Gamma(n+1)\Gamma^2(\Delta) \Gamma(2\Delta+2n-1)}\,.
\end{equation}
We will study corrections to the four point function in an expansion in powers of $1/N$
 \begin{equation}
 {\cal G}(u,v) =  {\cal G}^{(0)}(u,v)+ \frac{1}{N^2}  {\cal G}^{(1)}(u,v)+ \frac{1}{N^4}  {\cal G}^{(2)}(u,v) + \cdots
 \end{equation}
 The dimensions and OPE coefficients of double-trace operators will have a similar expansion
  \begin{eqnarray}
  \label{expansions}
 \Delta_{n,\ell} &=&\D_{n,\ell}^{(0)}+ \frac{1}{N^2} \gamma_{n,\ell}^{(1)}+  \frac{1}{N^4} \gamma_{n,\ell}^{(2)} + \cdots,\\
 a_{n,\ell} &=& a_{n,\ell}^{(0)}+ \frac{1}{N^2} a_{n,\ell}^{(1)} + \frac{1}{N^4} a_{n,\ell}^{(2)}+ \cdots
 \end{eqnarray}
Let us start by recalling the analysis at $O(1/N^2)$. Plugging the expansions for the dimensions and OPE coefficients into the conformal partial wave (CPW) decomposition (\ref{CPW}) we obtain
\begin{align} \label{forgone}
{\cal G}^{(1)}(u,v) = \sum_{n,\ \ell} u^{\Delta+n}  \left(a^{(1)}_{n,\ell} + \frac{1}{2} a^{(0)}_{n,\ell} \gamma^{(1)}_{n,\ell} \left(\log u+\frac{\partial}{\partial n}\right)\right) g_{2\Delta+2n+\ell,\ell}(u,v)~.
\end{align}
Due to the convergence properties of the OPE, the right-hand side displays explicitly the behaviour around $u=0$. On the other hand, to understand the behaviour around $v=0$ is more subtle. Each conformal block behaves as %
\e{}{g_{\D_p,\ell}(u,v)\big|_{v\rar 0} \sim {\tilde a}_{\D_p,\ell}(u,v) + {\tilde b}_{\D_p,\ell}(u,v) \log v\,,}
where ${\tilde a}_{\D_p,\ell}(u,v)$ and ${\tilde b}_{\D_p,\ell}(u,v)$ admit a series expansion around $u,v=0$. Hence each conformal block diverges logarithmically as $v \to 0$. However, infinite sums over the spin may generically change this behaviour. This will be important for us below.  

In \cite{Heemskerk:2009pn} a basis of solutions $\{\gamma_{n,\ell}^{(1)},a_{n,\ell}^{(1)} \}$ to the crossing equation \eqref{crossing} was constructed. Each of these solutions has support only for a bounded range of the spin $\ell$.  In this case, the analytic structure around both $u=0$ and $v=0$ is manifest, and the crossing equation $v^\Delta {\cal G}^{(1)}(u,v) = u^\Delta {\cal G}^{(1)}(v,u)$ can be split into different pieces, proportional to $\log u \log v$, $\log u$, $\log v$ and $1$ (times integer powers of $u$ and $v$). In \cite{Heemskerk:2009pn} it was argued that there is a one-to-one map between this basis of solutions to crossing and local four-point vertices in a bulk theory in AdS$_{d+1}$. Furthermore, let us mention that in Mellin space these solutions correspond simply to polynomials with appropriate symmetry properties. The degree of the polynomial determines for which range of spins the corrections $\{\gamma_{n,\ell}^{(1)},a_{n,\ell}^{(1)} \}$ are different from zero. Let us stress that these ``truncated" solutions are consistent with crossing only in the minimal set-up, in which only the identity and double-trace operators are present in the OPE ${\cal O} \times {\cal O}$. Later we will discuss what happens in more general cases. 

The aim of the present paper is to extend those solutions to consistent solutions to crossing at order $1/N^4$. We will assume the leading order solutions $\{\gamma_{n,\ell}^{(1)},a_{n,\ell}^{(1)} \}$ as given, and analyze consistency conditions on  $\{\gamma_{n,\ell}^{(2)},a_{n,\ell}^{(2)} \}$. Plugging the expansions \eqref{expansions} into the CPW decomposition (\ref{CPW}) we obtain 
\es{forgtwo}{{\cal G}^{(2)}(u,v) = \sum_{n,\ \ell} u^{\Delta+n}  &\Bigg(a^{(2)}_{n,\ell} + \frac{1}{2} a^{(0)}_{n,\ell} \gamma^{(2)}_{n,\ell} \left(\log u+\frac{\partial}{\partial n}\right)  \cr
& + \frac{1}{2}a^{(1)}_{n,\ell} \gamma^{(1)}_{n,\ell} \left(\log u+\frac{\partial}{\partial n}\right)\cr&+ \frac{1}{8}a^{(0)}_{n,\ell}  (\gamma^{(1)}_{n,\ell} )^2 \left(\log^2(u)  +2\log u \frac{\partial}{\partial n} + \frac{\partial^2}{\partial n^2} \right)  \Bigg)g_{2\Delta+2n+\ell,\ell}(u,v)\,.}
Note that the first line has the same structure as ${\cal G}^{(1)}(u,v)$, but with $\{\gamma_{n,\ell}^{(2)},a_{n,\ell}^{(2)} \}$ replacing $\{\gamma_{n,\ell}^{(1)},a_{n,\ell}^{(1)} \}$. The contribution from the other lines is uniquely fixed in terms of the solution at order $1/N^2$ and can be viewed as a source, or an inhomogeneous term, for the crossing equation
\begin{equation}
\label{crossingtwo}
v^\Delta {\cal G}^{(2)}(u,v) = u^\Delta {\cal G}^{(2)}(v,u)\,,
\end{equation}
interpreted as an equation for $\{\gamma_{n,\ell}^{(2)},a_{n,\ell}^{(2)} \}$. The analysis of this equation is much harder than the analysis at order $1/N^2$, since, as we will see momentarily, consistency with crossing implies that $\{\gamma_{n,\ell}^{(2)},a_{n,\ell}^{(2)} \}$ are different from zero for arbitrarily large spin. We will focus here on certain unambiguous contributions to the source terms, and understand their implications for the solution to the crossing equation.

\subsection{Implications from crossing at order $1/N^4$}
Let us focus on a specific contribution to ${\cal G}^{(2)}(u,v)$, which is the coefficient of $\log^2(u)$:
\begin{equation}
\left. {\cal G}^{(2)}(u,v) \right|_{\log^2(u) } = \sum_{n,\ell} u^{\Delta+n} \frac{1}{8}a^{(0)}_{n,\ell}  (\gamma^{(1)}_{n,\ell} )^2g_{2\Delta+2n+\ell,\ell}(u,v)\,.
\end{equation}
This contribution is unambiguously fixed in terms of the leading order solution. We can already make the following simple observation. Under crossing symmetry this term will map to a term with a divergence $\log^2 (v)$ as $v\to0$. Since each conformal block diverges at most logarithmically in this limit, such a contribution must come from an infinite sum over the spin, for a given twist. Hence it follows that the solution $\{\gamma_{n,\ell}^{(2)},a_{n,\ell}^{(2)} \}$ must be different from zero for arbitrarily large spins, even if the solution at order $1/N^2$ is truncated. From now on, it is convenient to restrict our considerations to truncated solutions at order $1/N^2$. More general solutions will be studied in Section \ref{infsupport}.

\subsubsection{Truncated solutions at order $1/N^2$}

If the solution at order $1/N^2$ truncates at spin $L$, we have:
\es{trunccase}{\left. {\cal G}^{(2)}(u,v) \right|_{\log^2(u) } &= \sum_{n} \sum_{\ell=0}^L u^{\Delta+n} \frac{1}{8}a^{(0)}_{n,\ell}  (\gamma^{(1)}_{n,\ell} )^2  g_{2\Delta+2n+\ell,\ell}(u,v)\\& \equiv u^\Delta \left( f(u,v) \log v+ g(u,v) \right)\,,}
where we have used the fact that the sum over spins truncates. $f(u,v)$ and $g(u,v)$ admit a series expansion in $u,v$ with integer powers, and can be computed in terms of the given leading order solution. As a consequence of crossing symmetry, $ {\cal G}^{(2)}(u,v)$ should also contain the following terms:
\begin{equation}
{\cal G}^{(2)}(u,v)  =  u^{\Delta}  \log^2 (v) \left(  f(v,u) \log u+ g(v,u)  \right)+ \cdots
\end{equation}
where the dots denote contributions proportional to $\log v$, or analytic at $v=0$. Given that the support of $\{\gamma_{n,\ell}^{(1)},a_{n,\ell}^{(1)} \}$ involves a finite range of the spin, the last two lines of (\ref{forgtwo}) cannot generate a $\log^2(v)$ behaviour, since each conformal block diverges at most logarithmically. Hence
\begin{equation}
\label{eqtruncated}
\left. \sum_{n,\ell} u^{n}  \frac{1}{2} a^{(0)}_{n,\ell} \gamma^{(2)}_{n,\ell} g_{2\Delta+2n+\ell,\ell}(u,v) \right|_{\log^2(v)}=  f(v,u)\,,
\end{equation}
and there is a similar equation involving the OPE coefficients $a_{n,\ell}^{(2)} $. (\ref{eqtruncated}) should be interpreted as an equation for $\gamma^{(2)}_{n,\ell}$, with the right-hand side $f(v,u)$ completely fixed in terms of the solution at order $1/N^2$. As already mentioned, since we need to reproduce an enhanced divergence on the left-hand side, we need to sum over an infinite number of spins. Furthermore, the divergence will arise from the region of large spin. In Section \ref{s4} we will adapt the algebraic method developed in \cite{Alday:2015ewa, Alday:2016njk} to determine the necessary large spin behaviour on $\gamma^{(2)}_{n,\ell}$ in order for (\ref{eqtruncated})  to be satisfied. The final answer is an expansion of the form
\begin{equation}
\label{crossingsol}
\gamma^{(2)}_{n,\ell}  = \frac{c_n^{(0)}}{\ell^{2\Delta}}\left( 1+ \frac{b_n^{(1)}}{\ell} +\frac{b_n^{(2)}}{\ell^2} + \cdots \right)\,,
\end{equation}
where all the coefficients of the expansion are actually computable. Hence we conclude that (\ref{eqtruncated}) actually fixes $\gamma^{(2)}_{n,\ell}$ up to solutions which decay faster than any power of the spin. Notice in particular that, from this point of view, we cannot expect to do any better, since there is always the freedom to add to any solution of \eqref{crossingtwo} a truncated solution which solves the homogeneous crossing equation (the same equation appearing at order $1/N^2$). In Section \ref{s5} we will study several examples. For these examples we will actually be able to do much more: we will be able to re-sum the whole series \eqref{crossingsol}, and extrapolate the results to finite spin.  

\subsubsection{Solutions with infinite support at order $1/N^2$}
\label{infsupport}

Let us now discuss the more general situation, in which the OPE of ${\cal O}$ with itself also includes single-trace operators. The two most important examples are ${\cal O}$ itself, of dimension $\Delta$, and the stress tensor $T_{\mu \nu}$, a spin two operator of dimension $\Delta_T=d$ and twist $\Delta_T-\ell=d-2$. These single-trace operators enter in the OPE decomposition with OPE coefficients squared of order $1/N^2$. In these cases ${\cal G}^{(1)}(u,v)$ contains the following terms
\begin{equation}
{\cal G}^{(1)}(u,v) = a_\Delta u^{\frac{\Delta}{2}} g_{\Delta,0}(u,v) + a_T u^{\frac{d-2}{2}} g_{d,2}(u,v) + \cdots
\end{equation}
Under crossing symmetry these map into terms of the form 
\begin{equation}
{\cal G}^{(1)}(u,v) \sim a_\Delta \frac{u^\Delta}{v^{\frac{\Delta}{2}}} \log u + a_T \frac{u^\Delta}{v^{\Delta-\frac{d-2}{2}}} \log u + \cdots\,,
\end{equation}
which lead \cite{Alday:2007mf,Komargodski:2012ek,Fitzpatrick:2012yx} to the following large spin behaviour for the anomalous dimensions of double-trace operators:
\begin{equation}
\label{particular}
\gamma^{(1)}_{n,\ell} \sim \frac{a_\Delta}{\ell^{\Delta}}\left(1+\cdots \right)+ \frac{a_T}{\ell^{d-2}} \left(1+\cdots \right)\,.
\end{equation}
This implies, in particular, that the leading order solution has infinite support in the spin. Single-trace operators can be seen as sources for the crossing equations, which are otherwise homogeneous. The general structure of the solution at order $1/N^2$ is then the sum of a solution to the equation with sources, with the behaviour (\ref{particular}), plus any of the truncated solutions studied above. Although any full-fledged conformal field theory contains the stress tensor, we will discuss its inclusion in a separate publication. In this paper, instead, we will consider only the presence of ${\cal O}$. This is relevant for correlators of $\phi^3$ theory on AdS. In this case
\begin{equation}
 \gamma^{(1)}_{n,\ell}  = a_\Delta  \gamma^{(1),\phi^3}_{n,\ell}  +\gamma^{(1),\rm trunc}_{n,\ell}  \,.
\end{equation}
where $\gamma^{(1),\phi^3}_{n,\ell}$ has support for all $\ell$, and $\gamma^{(1),\rm trunc}_{n,\ell}$ is any one of the truncated solutions.

As before, we can compute the piece proportional to $\log^2 (u)$ at order $1/N^4$. We obtain
\begin{equation}
\left. {\cal G}^{(2)}(u,v) \right|_{\log^2(u) } = \sum_{n,\ell} u^{\Delta+n} \frac{1}{8}a^{(0)}_{n,\ell}  (\gamma^{(1)}_{n,\ell} )^2  g_{2\Delta+2n+\ell,\ell}(u,v)\,,
\end{equation}
where now the sum over $\ell$ is not truncated. As we have already discussed, for a truncated solution the small $v$ behaviour is simply proportional to $\log v$, as for a single conformal block. In the case at hand, however, since the sum over the spin now does not truncate, we get an enhanced behaviour. More precisely, (\ref{particular}) leads to 
\begin{eqnarray}\label{huv}
\left. {\cal G}^{(2)}(u,v) \right|_{\log^2(u) } \sim a_\Delta^2 u^\Delta h(u,v) \log^2 (v) + \cdots \,.
\end{eqnarray}
Under crossing symmetry this contribution maps to itself, so that $h(u,v)=h(v,u)$. This is a consequence of crossing and the OPE expansion, and is completely independent of the new data $\{\gamma_{n,\ell}^{(2)},a_{n,\ell}^{(2)} \}$ at order $1/N^4$. In addition, as in \eqref{trunccase}, the sum above will contain contributions proportional to $\log v$
\begin{eqnarray}
\label{newv}
\left. {\cal G}^{(2)}(u,v) \right|_{\log^2(u) \log v} =u^{\Delta} f(u,v)\,.
\end{eqnarray}
However their computation is more subtle than before: one needs to perform the sum over the spin, and then expand for small $v$. Both the truncated and non-truncated parts of the solution will contribute to this term. The analysis of the crossing equations  is now more complicated. Under crossing the term \eqref{newv} maps to a term proportional to $\log u \log^2 (v)$. However, as the support of the solution at order $1/N^2$ is infinite, several terms in (\ref{forgtwo}) can produce an enhancement $\log^2 (v)$, and not only those involving $\gamma_{n,\ell}^{(2)}$. While this general case can also be analysed, note that the contributions from crossed terms to $( \gamma^{(1)}_{n,\ell} )^2 $, of the form $(2 a_\Delta  \gamma^{(1),\phi^3}_{n,\ell}  \times \gamma^{(1),\rm trunc}_{n,\ell})$ are much simpler to analyse. These crossed terms have a finite support, and their contribution to $\gamma_{n,\ell}^{(2)}$ can be computed exactly as explained above. We will discuss the interpretation of these contributions, and will compute them for specific examples, in Section \ref{s5}.

\section{Loop amplitudes in $AdS$}\label{s3}

The subleading solutions discussed in the previous section may be interpreted as one-loop contributions to correlation functions in AdS.
We now turn to constraining the general form of loop-level AdS amplitudes by studying features of the large $N$ expansion. We will employ the Mellin representation. One of the advantages of Mellin space is that AdS amplitudes have a transparent analytic structure as a function of the Mellin variables. This has been utilized in \cite{Penedones:2010ue} to write down compact and intuitive forms for tree-level Witten diagrams, and we will do the same here at one-loop. See \cite{0907.2407, Penedones:2010ue,Paulos:2011ie,Fitzpatrick:2011ia,Fitzpatrick:2011hu, Fitzpatrick:2011dm, Nandan:2011wc, Costa:2012cb,  Fitzpatrick:2012cg, Costa:2014kfa, Goncalves:2014rfa} for foundational work, and \cite{Goncalves:2014rfa,Alday:2015ota, Lowe:2016ucg,Nizami:2016jgt, Paulos:2016fap, Rastelli:2016nze, Gopakumar:2016wkt, Gopakumar:2016cpb} for some recent applications, of Mellin space in CFT. 

\ssec{Mellin amplitudes}

We now give a crash course in Mellin amplitudes in the context of the AdS/CFT correspondence. 

Consider the four-point function of identical operators $\la\O(x_1)\O(x_2)\O(x_3)\O(x_4)\ra$, related to an amplitude $\Gc(u,v)$ by \eqref{4pf}. By a double Mellin transform, we can trade $\Gc(u,v)$ for the Mellin amplitude, $M(s,t)$, defined to be
\e{amp}{\Gc(u,v) = {1\over (4\pi i)^2}\int_{-i\i}^{i\i} ds \,dt \,M(s,t)\,u^{t/2}v^{(\hat u-2\D)/2}\Gamma^2\left({2\D-t\over 2}\right)\Gamma^2\left({2\D-s\over 2}\right)\Gamma^2\left({2\D-\hat u\over 2}\right)\,,}
where ${\hat u} \equiv 4\Delta-s-t$.
The two integration contours run parallel to the imaginary axis, such that all poles of the gamma functions are on one side or the other of the contour.\footnote{The following formulae are specialized to the case of identical external operators, although many also hold for pairwise identical operators. A summary of the relevant formulae can be found in Appendix A of \cite{Costa:2012cb}. Their conventions are the same as in \eqr{amp} up to a shift $s_{\rm here} = s_{\rm there}+2\D$.} 
The product of gamma functions is totally symmetric in permutations of $(s,t,\hat u)$. Crossing symmetry of $\Gc(u,v)$ then implies total permutation symmetry of $M(s,t,\hat u)$:
\e{}{M(s,t) = M(s,\hat u) = M(t,s)\,.}

In a CFT with a weakly coupled AdS dual, the conformal block decomposition of $\Gc(u,v)$ translates into a sum of poles in $M(s,t)$. In a given channel, say the $t$-channel, the amplitude $M(s,t)$ has poles in $t$ at the twists of exchanged operators, and the residues encode the OPE coefficients:
\e{mope}{M(s,t) =\sum_{p}C_{\O\O\O_p}^2 \sum_{n=0}^\i {\Q_{\ell,n}(s;\t_p)\over t-(\t_p+2n)}\,,}
where the exchanged primary operator $\O_p$ has twist $\t_p=\Delta_p-\ell_p$. The pole at $t=\t_p+2n$ captures contributions of the twist-$(\t_p+2n)$ descendants of $\O_p$: schematically, these are the operators 
\e{}{(P^2)^n\O_p, ~~P_{\mu}(P^2)^n\O_p, ~~P_{\mu}P_{\nu}(P^2)^n\O_p,\ldots}
The residues  $\Q_{\ell,n}(s;\t_p)$ are the Mack polynomials, whose precise definition can be found in Appendix A of \cite{Costa:2012cb}. They have a spin index $\ell$ and a ``level'' $n$, and they depend on both the external and internal operator data. We will find it convenient to work with a ``reduced'' polynomial, $Q_{\ell,n}(s;\t)$, related to $\Q_{\ell,n}(s;\t)$ in general by \cite{Costa:2012cb}
\e{QQeq}{\mathcal{Q}_{\ell,n}(s;\t_p) =  Q_{\ell,n}( s;\t_p)\Bigg(-\frac{2\Gamma(\D_p+\ell) (\D_p-1)_\ell }{ 4^\ell\Gamma^4\left({\D_p+\ell\over 2}\right)n!(\D_p-{d\over 2}+1)_n  }
\frac{1 }{\Gamma^2\left( \frac{2\D -\t_p}{2}-n\right)}\Bigg)\,.}
For pairwise identical external operators as here, the $Q_{\ell,n}(s;\t_p)$ do not depend on the external dimensions. 

We will make use of the following facts about the $Q_{\ell,n}(s;\t_p)$. First, they are intimately related to the Mellin transform of the conformal blocks for exchange of a twist-$\t_p$ operator. In the lightcone expansion $u\ll 1$, the blocks take the form \eqr{Gblock} with
\e{}{g_{\D_p,\ell}(u,v) = \sum_{m=0}^{\i} u^m g^{(m)}_{\D_p,\ell}(v)\,.}
$g^{(0)}_{\D_p,\ell}(v) \equiv g_{\D_p,\ell}^\coll(v)$ is the collinear block,
\e{}{g^{\rm coll}_{\D_p,\ell}(v) = (1-v)^{\ell} \,{}_2F_1\left({\D_p+\ell\over 2},{\D_p+\ell\over 2},\D_p+\ell,1-v\right)~.}
The Mellin representation for all $g_{\D_p,\ell}^{(m)}(v)$ is
\es{gmv}{g^{(m)}_{\D_p,\ell}(v) &= \frac{2\Gamma(\D_p+\ell) (\D_p-1)_\ell }{ 2^\ell\Gamma^4\left({\D_p+\ell\over 2}\right)m!(\D_p-{d\over 2}+1)_m  }\\&\times\int_{-i\i}^{i\i} {ds\over 8\pi i}v^{-(s+\t_p-2\D)/2}Q_{\ell,m}(s-\t_p;\t_p) \Gamma^2\left({2\D-s\over 2}\right)\Gamma^2\left({s\over 2}\right)\,.}
Second, $Q_{\ell,0}(s;\t_p)$ takes the explicit form
\e{}{Q_{\ell,0}( s;\t_p)= \frac{2^{\ell}\left(\frac{\t_p}{2}\right)^2_{\ell}}{
 \left(\t_p+\ell-1\right)_{\ell}} \,_3F_2\!\left(-\ell,\t_p+\ell-1,\frac{-s}{2};\frac{\t_p}{2}, \frac{\t_p }{2};1\right)\,.
   \label{QJ0} }
This has the useful property that
\e{}{Q_{\ell,0}(s-\t_p;\t_p) = (-1)^\ell Q_{\ell,0}(-s;\t_p)~, \quad \ell\in\mathbb{Z}\,.}
These obey an orthogonality relation \cite{Costa:2012cb}, which can be written\footnote{Analogous orthogonality relations exist at higher $n$ but have not been calculated explicitly, to the best of our knowledge.}
\e{ortho}{\int_{-i\infty}^{i\infty}\frac{ds}{4\pi i}Q_{\ell,0}(-s;\t_p)Q_{\ell',0}(-s;\t_p)\Gamma^2\left(\frac{s}{2}\right)\Gamma^2\left(\frac{\t_p-s}{2}\right)={ 4^{\ell}\ell!\, \Gamma^4 \left(\frac{\t_p}{2}\right)\Gamma^2\left({2\D-\t_p\over 2}\right)\over \Gamma(\D_p+\ell)(\D_p-1)_{\ell}}\d_{\ell,\ell'}~.}
Given some amplitude expanded in the lightcone regime of small $u$ and fixed $v$, this relation allows one to strip off the coefficient of the leading-twist, spin-$\ell$ lightcone block $g^\coll_{2\D+\ell,\ell}(v)$ due to the exchange of $\OO_{0,\ell}$.

We now develop the AdS loop expansion of the connected piece of $\Gc(u,v)$ and $M(s,t)$, corresponding to the $1/N$ expansion of some holographic CFT:
\e{}{\Gc(u,v) = {1\over N^2}\Gc_{\rm tree}(u,v) + {1\over N^4}\, \Gc_{\rm 1-loop}(u,v) + \cdots}
In the language of Section \ref{crossingloops}, $\Gc_{\rm tree} = \Gc^{(1)}$ and $\Gc_{\rm 1-loop} = \Gc^{(2)}$. Likewise for the Mellin amplitude,
\e{}{M(s,t) = {1\over N^2} M_{\rm tree}(s,t) +  {1\over N^4}\, M_{\rm 1-loop}(s,t) + \cdots}
To set the stage for $\mloop$, we need to review the structure of $\mtree$. 

\ssec{Tree-level}

We are interested in paradigmatic large $N$ holographic CFTs which have a large gap in their spectra, or generalized free field sectors thereof. These are dual to weakly coupled gravity, coupled to a finite number of light fields. The spectra of these theories consist of ``single-trace'' operators $ \O_i$ and their ``multi-trace'' composites $ [\O_i\O_j],[\O_i\O_j\O_k]$, etc., that are dual to single-particle and multi-particle states in the bulk, respectively. As discussed above, the CFT conformal block decomposition of $\Gc_{\rm tree}$ only includes single-trace and double-trace exchanges. 

There are two salient points about $\mtree$. The first is that its only poles come from the single-trace exchanges of $\Gc_{\rm tree}$. These each contribute as in \eqr{mope}. The second is that the double-trace exchanges of $\Gc_{\rm tree}$ are accounted for by the explicit $\Gamma^2$ factors in the Mellin integrand \eqref{amp}, one for each channel, which have double poles at $\t=2\D+2n$. This makes explicit a fact about holographic CFTs: at tree-level, the single-trace OPE data completely determine the double-trace OPE data, up to the presence of regular terms in $M_{\rm tree}$. 

The gamma function residues include a $\log u$ term and a term regular at small $u$,
\e{treeres}{\Res{t=2\D+2n} \Gc_{\tree}(u,v)  = u^{\D+n}\big(A_n(v)\log u + B_n(v)\big)\,, }
where
\e{}{A_n(v) ={1\over 4\pi in!^2} \int_{-i\i}^{i\i} ds\, v^{-{(s+2n)\over 2}}M_{\rm tree}(s,2\D+2n)\G^2\left({s+2n\over 2}\right)\G^2\left({2\D-s\over 2}\right)\,.}
$B_n$ may be extracted similarly. Matching this to \eqr{forgone}, one can extract $\g^{(1)}_{n,\ell}$ and $a^{\1}_{n,\ell}$ by picking off the contribution proportional to the appropriate conformal block in the $u\ll1$ expansion. The $A_n\log u$ terms contain $\g^{\1}_{n,\ell}$, and the $B_n$ terms contain $a^{(1)}_{n,\ell}$.  The extraction of the leading-twist double-trace operator data, like $\g^{\1}_{0,\ell}$, is especially simple: from \eqr{forgtwo}, we require
\e{}{A_0(v) = \frac{1}{2} \sum_{\ell~{\rm even}}  a^{(0)}_{0,\ell} \gamma^{(1)}_{0,\ell}\, g^\coll_{2\Delta+\ell,\ell}(v)\,.}
The Mellin representation of $g_{2\D+\ell,\ell}^\coll(v)$ may be written
\e{gcoll}{g_{2\D+\ell,\ell}^\coll(v) = (-1)^{\ell}16 \,d_{\D,\ell}\int_{-i\i}^{i\i} {ds\over 8\pi i}v^{-s/2}Q_{\ell,0}(-s;2\D)\G^2\left({s\over 2}\right)\G^2\left({2\D-s\over 2}\right)\,,}
where we have introduced a convenient combination for future use,
\e{cdl}{d_{\D,\ell} \equiv \frac{\Gamma(2\Delta+2\ell) (2\D+\ell-1)_\ell }{ 2^{\ell +3}\Gamma^4\!\left( {\Delta +\ell}\right)}~.}
Upon using the orthogonality relation \eqr{ortho}, one finds the explicit formula \cite{Goncalves:2014rfa}
\e{g1form}{\g^{\1}_{0,\ell} = -{1\over 2\pi i}\int_{-i\i}^{i\i} ds\, M_{\tree}(s,2\D)\G^2\left({s\over 2}\right)\G^2\left({2\D-s\over 2}\right){}_3F_2(-\ell,\ell+2\D-1,{s\over 2};\D,\D;1)\,.}
A similar analysis allows one to extract $a^{\1}_{0,\ell}$. For higher $n$, one must deconvolve the subleading corrections $g_{\D,\ell}^{(m)}(v)$ to the small $u$ blocks, from the leading contributions coming from $n>0$ double-trace primaries. Expressions and an algorithm for computing $g^{(m)}_{\D,\ell}(v)$ can be found in \cite{Alday:2015ewa}.

\ssec{One-loop}
We now turn to $\mloop$. In a general CFT, this may receive various contributions. These fall into two categories: 

First, there are loop corrections to tree-level data. This includes mass, vertex and wave function renormalization of fields already appearing at tree-level; that is, $O(1/N^4)$ changes to the norms, dimensions and OPE coefficients of CFT operators appearing in the planar correlator. Corrections to the OPE data of single-trace operators can arise, but they can be easily taken into account by expanding the leading order solutions, and we will assume for simplicity that they vanish. Note that in any case these cannot be determined by the crossing equations, which have solutions for any such data.

Second, as discussed in Appendix \ref{crossing_ops}, there are new operator exchanges that do not appear at tree-level, due to large $N$ factorization. A simple example in a theory of gravity coupled to a scalar field is the appearance of two-graviton intermediate states, dual to $[TT]$-type double-trace operators, in the scalar correlator $\la \O\O\O\O\ra$. 

A universal contribution in any holographic CFT is the next-order correction to the tree-level $[\O\O]_{n,\ell}$ OPE data, namely, $\g^{\2}_{n,\ell}$ and $a^{\2}_{n,\ell}$. Let us write the double-trace piece of the total one-loop amplitude as
\e{}{ \mloop^{\OO}(s,t)~.}
We note that in simple AdS effective theories like $\l\phi^4$ dressed with any number of derivatives, this is the {\it full} amplitude. More precisely, for any theory in which no single-trace operators appear in the OPE (dual to theories in AdS with no cubic vertices), and in which there are no extra double-trace operators in the OPE (dual to the absence of four-point couplings to other fields in AdS), we have
\e{}{\mloop(s,t) = \mloop^{\OO}(s,t).}
When $\lbrace \t_i\rbrace\in2\mathbb{Z}$ in more general theories, there are similar simplifications, as we discuss in Section \ref{s7}.

We now establish the following simple but powerful claim: all poles and residues of $\mloop^{\OO}$ are completely fixed by tree-level data. It follows that $\g^{\2}_{n,\ell}$ and $a^{\2}_{n,\ell}$ are  fixed by $\g^{\1}_{n,\ell}$ and $a^{\1}_{n,\ell}$. 

Recall that the contribution of $[\O\O]_{n,\ell}$ to $\Gc^{\2}$ takes the form given in \eqr{forgtwo}:
\e{50}{{\cal G}^{(2)}(u,v)= \sum_{n,\ell} u^{\Delta+n} \left(\frac{\log^2 (u)}{8}a^{(0)}_{n,\ell}  (\gamma^{(1)}_{n,\ell} )^2   g_{2\Delta+2n+\ell,\ell}(u,v) + O(\log u)\right)\,.}
The point is that there is a $\log^2 (u)$ term whose coefficient is completely fixed by tree-level data. In order to correctly produce this term at each power $u^{\D+n}$ ($n=0,1,2,\ldots$), two things must happen: 

{\bf 1)} $\mloop$ must acquire simple poles at $\t=2\D+2n$ for $n=0,1,2,\ldots$.

{\bf 2)} The residues are fixed by $\g^{\1}_{n,\ell}$ so as to match \eqr{50}. 

This is true in each of the $s,t,\hat u$ channels, so we can focus on just one, and trivially add the crossed channels to get the full $\mloop^{\OO}$. Showing the $t$-channel for concreteness, we have thus determined that
\e{mloopgen}{\mloop^{\OO}(s,t) = \sum_{n=0}^\i {R_n(s)\over t-(2\D+2n)} + f_{\reg}(s,t) + ({\rm crossed})}
for some residues $R_n(s)$. This argument does not determine any possible regular terms in $\mloop^{\OO}$, so we have allowed for a function $f_{\reg}$. We drop this for now, but will return to it shortly; as we will see, $f_{\reg}$ is not unique. 

To determine the residues $R_n(s)$, we use the same technique as at tree-level. We have
\e{abc}{\Res{t=2\D+2n} [\Gc_{\rm 1-loop}(u,v)]  = u^{\D+n}\big(A_n(v)\log^2 (u) + B_n(v)\log u +  C_n(v)\big)\,, }
where $A_n,B_n,C_n$ are easily determined by plugging $\mloop$ of \eqr{mloopgen} into the Mellin amplitude formula \eqr{amp}. To fix the $R_n(s)$ we insist upon equality of $A_n$ with the $\log^2(u)$ term in \eqr{50}. Given the Mellin representation \eqr{gmv} of the conformal blocks in the $u\ll1$ expansion, this fixes the $R_n(s)$ completely for every $n$. 

For example, the leading residue $R_0(s)$ is determined by the following equation:
\e{}{A_0(v) = {1\over 4\pi i}\int_{-i\infty}^{i\infty} ds \,R_0(s)v^{-s/2}\Gamma^2\left({2\D-s\over 2}\right)\Gamma^2\left({s\over 2}\right) = {1\over 8}\sum_{\ell~{\rm even}}a_{0,\ell}^{(0)}(\g^{(1)}_{0,\ell})^2 g_{2\D+\ell,\ell}^\coll(v)\,.}
Using the Mellin representation \eqr{gcoll} of $g^\coll_{2\D+\ell,\ell}(v)$ determines $R_0(s)$ to be 
\e{R0s}{R_0(s) = \sum_{\ell=0}^\i a_{0,\ell}^{(0)}(\g^{(1)}_{0,\ell})^2 d_{\D,\ell}\,Q_{\ell,0}(-s;2\D)\,,}
where $d_{\D,\ell}$ was defined in \eqr{cdl}, and $Q_{\ell,0}(-s;2\D)$ is the polynomial \eqr{QJ0} at intermediate twist $2\D$. Note that in the formula for $R_0(s)$, the coefficients of $Q_{\ell,0}(-s;2\D)$ are manifestly positive. Higher $R_n(s)$ can, with some work, be extracted similarly. By matching $B_n$ in \eqr{abc} to the $\log u$ terms in \eqr{forgtwo}, one can compute $\g^{\2}_{n,\ell}$, as we will show in an explicit example shortly.

\sssec{UV divergences and $f_{\reg}$}\label{s331}
We now return to the physics of the function $f_{\reg}$ in \eqr{mloopgen}. 

The first point to note is that \eqr{mloopgen} is a solution to crossing for {\it any} permutation-symmetric $f_{\reg}$. The minimal solution is $f_{\reg}=0$. Indeed, $f_{\reg}$ reflects the freedom to add a homogeneous solution to the second-order crossing equations \eqr{forgtwo}. Such solutions sit in one-to-one correspondence with quartic contact interactions in AdS; in Mellin space, these are simply crossing-symmetric polynomial amplitudes \cite{Heemskerk:2009pn, Penedones:2010ue}. So we should think of $f_{\reg}$ as a choice of one-loop renormalization conditions for the quartic part of the effective action for the light fields in AdS, dual to a choice of one of the infinite solutions to the one-loop crossing equations that differ by polynomials, i.e. finite local counterterms in AdS. 

What happens when the bulk theory is one-loop divergent? In this case, one must include in the bulk some diverging local counterterms to restore finiteness. Due to their locality, these again appear in the function $f_{\reg}$. This was explained in general terms in the Introduction;  for more discussion, see Appendix \ref{UVdiv}. In the explicit results for scalar theories that will follow in Section \ref{s5}, we will see very nicely in detail how bulk UV divergences show up in the one-loop CFT correlators.

For all of these reasons, $f_{\reg}$ is not unique, and may sometimes not be finite before renormalizing the bulk theory. We note that various high-energy limits, such as the Regge limit of large $s$ and fixed $t<0$, may place some constraints on $f_{\reg}$, see ${\it e.g.}$ \cite{Alday:2016htq}.

\sssec{Examples}
Let us treat some simple and instructive examples. 

The first is $\phi^4$ theory in AdS. There is a single non-trivial one-loop diagram, the bubble diagram, in each channel. (There are also diagrams which lead to mass and wave function renormalization of $\phi$, but these only serve to renormalize $\mtree$, which is anyway constant in this case.) On the CFT side, as explained earlier in Section \ref{crossingloops}, there are {\it only} double-trace operators in the one-loop four-point function. So, up to an additive constant,
\e{}{\mloop(s,t) = \mloop^{\OO}(s,t) = \sum_{n=0}^\i R_n\left({1\over s-(2\D+2n)}+ {1\over t-(2\D+2n)}+ {1\over \hat u-(2\D+2n)}\right)}
is a solution to the one-loop crossing problem, with {\it constant} residues $R_n$, because $\g^{\1}_{n,\ell}\propto \delta_{\ell,0}$. From \eqr{R0s}, 
\e{R0phi4}{R_0 =  \frac{\Gamma(2\Delta) }{ 8\Gamma^4\!\left( {\Delta }\right)}\, a_{0,0}^{(0)}(\g^{(1)}_{0,0})^2\,.}
Moreover, it is a simple matter to extract $\g^{\2}_{0,\ell>0}$. For $\ell>0$, the one-loop OPE data are the leading correction to the mean field theory OPE data. From \eqr{forgtwo}, we see that
\e{59}{\Gc^{(2)}(u,v)\Big|_{\OO_{0,\ell>0}} = u^\D\left(a^{(2)}_{0,\ell}+\frac{1}{2} a^{(0)}_{0,\ell} \gamma^{(2)}_{0,\ell}  \left(\log u+\frac{\partial}{\partial n}\right)\right)g_{2\Delta+2n+\ell,\ell}(u,v)\big|_{n=0}\,.}
This takes the same form as the first-order crossing equations \eqr{forgone}. So by the same logic that led to \eqr{g1form}, we have
\e{g2phi4}{\g^{\2}_{0,\ell>0} = -{1\over 2\pi i}\int_{-i\i}^{i\i} ds\,\mloop'(s,2\D)\G^2\left({s\over 2}\right)\G^2\left({2\D-s\over 2}\right){}_3F_2(-\ell,\ell+2\D-1,{s\over 2};\D,\D;1)\,.}
where $\mloop'(s,2\D)$ is defined as $\mloop(s,2\D)$ minus the pole at $t=2\D$.\footnote{The $\log u$ term coming from the triple pole at $t=2\D$ only contributes to $\ell=0$. $\mloop'(s,2\D)$ is also equivalent to just keeping the $s$- and $\hat u$-channels.} The formula \eqr{g2phi4} extends to any theory with a truncated spectrum, $\g^{\1}_{n,\ell>L}=0$, of the sort considered in \cite{Heemskerk:2009pn}: simply replace $\g^{\2}_{0,\ell>0}$ with $\g^{\2}_{0,\ell>L}$.

A more interesting example is $\phi^3$ theory. At tree-level, $\mtree$ includes $\phi$ exchange alone,
\e{}{M_{\tree}(s,t) = C_{\O\O\O}^2\sum_{n=0}^\i {\Q_{0,n}(s;\D)\over t-(\D+2n)} + (\text{crossed})}
where $\phi$ is dual to a dimension $\D$ scalar operator $\O$. There is no regular term \cite{Penedones:2010ue, Fitzpatrick:2011ia, Paulos:2011ie}. At one-loop, there is a single non-trivial diagram in each channel, the scalar box diagram shown in Figure \ref{loopexp}. There is also renormalization of $\mtree$, which generally requires also adding the contribution of a bulk $\phi^4$ term. The only operators appearing in the OPE up to order $1/N^4$ are $\cO$ and $[\cO \cO]_{n,\ell}$. Denoting the parts of the one-loop amplitude that simply renormalize the tree-level CFT data as $\widehat{M}_{\rm tree}$, the full one-loop amplitude is
\e{m70}{\mloop(s,t) = \widehat{M}_{\rm tree}(s,t) + \sum_{n=0}^\i\left({R_n(s)\over t-(2\D+2n)}+ {(\text{crossed})}\right)\,.}
We can explicitly calculate the leading residue $R_0(s)$ from \eqr{R0s}. To be concrete, let us take $\D=2=d$. First we need to know $\g^{\1}_{0,\ell}$. One can compute, either using $\mtree$ and \eqr{g1form} or a spacetime decomposition of $\Gc^{\1}$,
\e{phi3g10l}{\g^{\1}_{0,\ell>0} = -{2C_{\O\O\O}^2\over(\ell+1)(\ell+2)}~,\quad \g^{\1}_{0,0} = -{5C_{\O\O\O}^2\over6}~.}
We prove this in Section \ref{s6} and Appendix \ref{appd}.  Note that $\g^{\1}_{0,\ell\rar\i} \sim -2C_{\O\O\O}^2\ell^{-2}$, consistent with the lightcone bootstrap \cite{Komargodski:2012ek,Fitzpatrick:2012yx} (cf. \eqr{lcboot3}). Then
\begin{equation}
R_0(s)=\left(\frac{25}{12}+\sum_{\ell=2}^{\infty}\frac{(3+2\ell)}{(1+\ell)(2+\ell)}{}_3F_2(-\ell,3+\ell,\frac{s}{2}; 2,2;1)\right)C_{\O\O\O}^4~.
\end{equation}
This gives us a piece of the one-loop box diagram of the $\phi^3$ theory on AdS; this diagram has not yet been computed.

\sssec{Large spin and the lightcone bootstrap}

One reason that knowing the leading residue $R_0(s)$ is useful is because, as we now establish, in the large spin limit $\ell\rar\i$, $\g^{\2}_{0,\ell}$ is controlled by the leading pole of $\mloop$. 

Consider the formula \eqr{g2phi4}, applicable to the truncated theories. At large spin, ${}_3F_2$ has two branches of power series:
\e{3f2asy}{ {}_3F_2(-\ell,\ell+2\D-1,{s\over 2};\D,\D;1) \approx {\G^2(\D)\over \G^2({2\D-s\over 2})}\beta_{J}(s) + {(-1)^{\ell}\G^2(\D)\over \G^2({s\over 2})} \beta_{J}(2\D-s)~,}
where $\beta_{J}(s)$ is the expansion for large spin. We have introduced the ``conformal spin,''
\e{jsq}{J^2 = (\ell+\D)(\ell+\D-1)~,}
because $\beta_{J}(s)$ actually has an expansion in inverse powers of $J^2$. This will be useful in Sections \ref{s4} and \ref{s5}. The first terms of this expansion are
\begin{equation}\label{3f2asy2}
\beta_{J}(s)=J^{-s}\left(1+\frac{3 \Delta s^2-s^3-3 s^2-2 s}{12 J^2}+\dots\right)~.
\end{equation}
At leading order in large $\ell$, we have, from the $s$-channel,\footnote{Recall that $\ell\in 2\mathbb{Z}$.}
\e{g2large}{\g^{\2}_{0,\ell\gg 1}\Big|_{\rm s-channel} = -{{\G^2(\D)}\over 2\pi i}\int_{-i\i}^{i\i} ds\, \left({R_0(2\D)\over s-2\D}+\ldots\right) \left(\G^2\left({s\over 2}\right)\ell^{-s} + \G^2\left({2\D-s\over 2}\right) \ell^{s-2\D}\right)\,.}
To evaluate the first term, we close the contour to the right, picking up the leading pole at $s=2\D$. To evaluate the second term, we close the contour to the left where there are no poles, thus yielding zero. The $\hat u$-channel contributes identically, due to symmetry of the Mellin integrand (the role of the two terms is reversed). In all, we arrive at
\e{g2large2}{\g^{\2}_{0,\ell\gg 1} \sim -{2\G^4(\D)}{R_0(2\D)\ell^{-2\D}}\,.}
This represents the contribution to $\g^{\2}_{0,\ell\gg1}$ from the double-trace piece of $\mloop$. 

For example, in $\phi^4$ theory, using \eqr{R0phi4} one finds 
\e{phi4large}{\g^{\2}_{0,\ell\gg1} \sim -{a_{0,0}^{\0}(\g^{\1}_{0,0})^2\Gamma(2\D)\over 4}\, \ell^{-2\D}\,.}
This precisely matches the result from the lightcone bootstrap, in an interesting way. The general formula for large spin asymptotics of double-trace anomalous dimensions is \cite{Komargodski:2012ek,Fitzpatrick:2012yx}
\e{lcboot3}{\g_{0,\ell\gg 1} \sim -{c_{\t_m}\over \ell^{\t_m}}~, \quad c_{\t_m} = {C^2_{\O\O\O_{m}}\over N_{\O\O}^2 N_{\O_m\O_m}} \left({2\G^2(\D)\G(\t_m+2\ell_m)\over \G^2(\D-{\t_m\over 2})\G^2({\t_m\over 2}+\ell_m)}\right)\,,}
where $\O_m$ is the minimal twist operator appearing in the $\O\times\O$ OPE, of twist $\t_m$ and spin $\ell_m$, and $N_{\O\O}, N_{\O_m\O_m}$ are norms. In the case of a $\phi^4$ generalized free field, the leading twist operators are themselves the double-trace operators $\OO_{0,\ell}$; this explains why the leading asymptotic is a one-loop effect. At first sight, we need to sum over all $\ell$, introducing subtleties discussed in e.g. Section 4.3 of \cite{Komargodski:2012ek}. However, working in a $1/N$ expansion, the minimal twist operator is the $\OO_{0,0}$ operator only, because $\g^{\1}_{0,\ell>0}=0$. So, we should plug in
\es{pvals}{\t_m = 2\D+\g^{\1}_{0,0}+\ldots~, \quad \ell_m=0~, \quad C_{\O\O\O_m}^2 = a^{\0}_{0,0}~, \quad N_{\O\O}=1~, \quad N_{\O_m\O_m} = 2\,,}
and expand to leading order in $\g^{\1}_{0,0}$. This precisely reproduces \eqr{phi4large}, providing a nice check on our formula \eqr{R0s}.

\ssec{Enter crossing}\label{s3cross}
So far, we have shown how $\mloop^{\OO}$, and hence the one-loop OPE data $\g^{\2}_{n,\ell}$ and $a^{\2}_{n,\ell}$, can be fixed up to regular terms by large $N$ considerations alone. In practice, doing so beyond the leading residue is complicated, and seems to require knowing $\g^{\1}_{n,\ell}$ for all $n$. 

We now advocate a different approach that uses crossing symmetry: namely, we solve for the one-loop OPE data $\g^{\2}_{n,\ell}$ and $a^{\2}_{n,\ell}$, from which we can construct $\mloop$. As a paradigm for how to actively reconstruct $\mloop$, we turn to AdS $\phi^4$ theory. From \eqr{g2phi4}, we make the following observation. If we know $\g^{\2}_{0,\ell}$ in the large spin expansion, then by expanding the integrand at large spin, we can reconstruct all residues of $\mloop$, pole by pole! Note that in this case, neither $\g^{\2}_{n>0,\ell}$ nor $a^{\2}_{n,\ell}$ are required. In truncated theories where tree-level anomalous dimensions are generated only up to spin $L$, only a subset of the $\g^{\2}_{n,\ell}$ data is needed.\footnote{When $L=0$, the residues of $\mloop$ are constants, so it takes only one value of $n$ to reconstruct these constants; for a theory truncated at spin $L$, the residues are degree-$L$ polynomials -- see e.g. \eqr{R0s} -- so the large spin expansion of only $L+1$ pieces of data are needed for full reconstruction. Note also that the $a^{\2}_{n,\ell}$ may, but need not, be used. This implies that there is a one-loop analog of the tree-level derivative relation, $a^{\1}_{n,\ell} = {1\over 2}\p_n(a^{\0}_{n,\ell}\g^{\1}_{n,\ell})$, discovered in \cite{Heemskerk:2009pn} and proven in \cite{Fitzpatrick:2011dm}. For $\ell>L$, this relation holds at one-loop too, since the $O(1/N^4)$ crossing equations are identical to the $O(1/N^2)$ equations, as discussed around \eqr{59}. For $\ell\leq L$, and for totally general holographic theories, it would be very interesting to derive its analog, though we will not do so here.}

\sec{Solving the one-loop crossing equations}\label{s4}
In this section we introduce a machinery to solve the one-loop crossing equations for a generic large $N$ CFT. In the next section, we apply this to explicit examples of scalar theories in AdS, dual to generalized free field sectors of holographic CFTs. 

In Section \ref{crossingloops} we have obtained equation \eqref{eqtruncated} for $\gamma_{n,\ell}^{(2)}$ as a consequence of crossing symmetry. Since each conformal block diverges at most logarithmically as $v\to 0$, we need to sum an infinite number of them in order to obtain the enhanced divergence $\log^2 (v)$. As we will show below, this fixes uniquely the behaviour of $\gamma_{n,\ell}^{(2)}$ as a series expansion in $1/\ell$. For simplicity let us focus on $\gamma_{0,\ell}^{(2)}$. We have
\begin{equation}
\label{gammaeq}
\sum_{\ell ~\text{even}} \frac{1}{2} a_{0,\ell}^{(0)}\gamma_{0,\ell}^{(2)} g_{2\Delta+\ell,\ell}^\coll(v) = f(v) \log^2 (v) + \cdots
\end{equation}
where $g_{2\Delta+\ell,\ell}^\coll(v)$ is the small $u$ limit of the full conformal block, and the dots denote terms whose divergence is not enhanced with respect to a single conformal block. Our task is to find $\gamma_{0,\ell}^{(2)}$ as a series expansion in $1/\ell$ for a given $f(v)$. In the following we introduce a method to solve this problem. 

\subsection{The Casimir method}
\label{casimir}
\subsubsection{Basic Idea}

As just mentioned, our task is to find $\gamma_{0,\ell}^{(2)}$ as a series expansion in $1/\ell$ such that (\ref{gammaeq}) holds. In order to do this the following property will be important. There exists a Casimir operator such that
\begin{equation}
\label{Casimir}
{\cal C} g_{2\Delta+\ell,\ell}^\coll(v)=J^2g_{2\Delta+\ell,\ell}^\coll(v)\,,
\end{equation}
where the collinear Casimir and eigenvalue are given by
\begin{eqnarray}
{\cal C} &=& v(1-v)^2 \partial^2_v +(1-v)(1-v-2v \Delta)\partial_v +\Delta(v \Delta-1)  \,,\\
J^2 &=& (\ell+\Delta)(\ell+\Delta-1)\,.
\end{eqnarray}
Notice that acting repeatedly with the Casimir operator on $v^n \log^2(v)$ we will produce a negative power of $v$ after a finite number of times. This is not true for $v^n \log v$. The idea is then very simple: acting with the Casimir operator on both sides of (\ref{gammaeq}), we increase the degree of the large $\ell$ divergence, and we are able to explore more and more terms in the $1/\ell$ expansion. In the following we show how to convert this into an algebraic problem. 

We will follow the same strategy as in \cite{Alday:2015ewa, Alday:2016njk} adapted to the specific problem at hand. We start by considering the sum \eqref{gammaeq} without the insertion of the anomalous dimension
\begin{equation}
\label{normalisation}
\sum_{\ell ~\text{even}}  a_{0,\ell}^{(0)} g_{2\Delta+\ell,\ell}^\coll(v) = \frac{1}{v^\Delta}+ \cdots
\end{equation}
where the dots denote finite terms as $v \to 0$. Next we consider spin-dependent insertions proportional to negative powers of $J^2$.  It turns out that insertions of the form $J^{-2\Delta-2n}$ will generate an enhanced divergence $\log^2(v)$ \footnote{This can be understood as follows. For small values of $v$ the sum over the spin in (\ref{normalisation}) peaks at $\ell \sim 1/\sqrt{v}$. Hence an extra insertion of $J^{2m}$ will lead to a power law divergence $v^{\Delta-m}$. For $m \to \Delta$ the power law behaviour disappears, and we obtain a $\log^2(v)$ divergence.}. It is then convenient to introduce the following basis of functions
\begin{equation}
\label{hdef}
\sum_{\ell ~\text{even}} \left(\frac{1}{J^2} \right)^{\Delta+n}  a_{0,\ell}^{(0)} g_{2\Delta+\ell,\ell}^\coll(v) = h^{(n)}(v) \log^2(v)+\cdots
\end{equation}
where the dots denote terms whose divergence as $v\to 0$ is not enhanced with respect to that of a single conformal block.  Assuming the following expansion for $\gamma_{0,J}^{(2)}$
\begin{equation}
\gamma_{0,J}^{(2)} = \frac{2}{J^{2\Delta}} \left(b_0 + \frac{b_1}{J^2} + \frac{b_2}{J^4} + \frac{b_3}{J^6}+\cdots \right)\,,
\end{equation}
we see that (\ref{gammaeq}) is equivalent to
\begin{equation}
b_0 h^{(0)}(v) + b_1 h^{(1)}(v) + b_2 h^{(2)}(v) + \cdots  = f(v)~.
\end{equation}
So, once the basis $h^{(n)}(v)$ is found, finding the $b_i$ is equivalent to the problem of writing the given $f(v)$ in such a basis. Furthermore, it turns out $h^{(n)}(v) \sim v^n$ for small $v$, so that the equation above can be solved order by order in powers of $v$ and is completely algebraic. 

\subsubsection{The basis $h^{(n)}(v)$}
The action of the Casimir operator (\ref{Casimir}) on collinear conformal blocks translates into a recurrence relation for the sequence of functions $h^{(n)}(v)$
\begin{equation}
{\cal C} h^{(n)}(v) = h^{(n-1)}(v)\,.
\end{equation}
For integer $\Delta$ we can take the defining relation for $h^{(0)}(v)$ to be
\begin{equation}
{\cal C}^\Delta (h^{(0)}(v) \log^2 (v)) = \frac{1}{v^\Delta} + \cdots
\end{equation}
where the dots denote a contribution which is not enhanced with respect to a single conformal block. In the following we will find it convenient to define
\begin{equation}
{\cal C}= (1-v)^{-\Delta} \hat {\cal C} (1-v)^\Delta,~~~~h^{(n)}(v)  = \frac{\hat h^{(n)}(v) }{(1-v)^\Delta}\,,
\end{equation}
and to change variables to 
\e{}{\zeta = \frac{v}{1-v}~.}
In these variables the problem is equivalent to
\begin{equation}
\label{basiscond}
 \hat {\cal C}^\Delta (\hat h^{(0)}(\zeta) \log^2(\zeta)) = \frac{1}{\zeta^\Delta} + \cdots,~~~~  \hat {\cal C} \hat h^{(n)}(\zeta) =  \hat h^{(n-1)}(\zeta),
\end{equation}
where dots denote terms which are not enhanced respect to $\log (\zeta)$, and $ \hat {\cal C} $ takes a very simple form:
\begin{equation}
  \hat {\cal C}=(1+2\zeta) \partial_\zeta+\zeta(1+\zeta)\partial^2_\zeta\,.
\end{equation}
The conditions (\ref{basiscond}) have to be supplemented by the behaviour of $\hat h^{(n)}(\zeta)$ around $\zeta=0$. We require $\hat h^{(n)}(\zeta)$ to admit a series expansion around $\zeta=0$, with integer powers and starting with $\hat h^{(n)}(\zeta) \sim \zeta^n$. This fixes the solutions uniquely. Let us mention that the absence of $\log^2(\zeta)$ terms in the first relation of (\ref{basiscond}) also implies the necessary condition $ \hat {\cal C}^\Delta (\hat h^{(0)}(\zeta)) = 0$. A more careful analysis actually shows that for $\Delta >1$ and for the solution with the correct boundary conditions a slightly stronger equation is satisfied
\begin{equation}
\label{neccond}
 \hat {\cal C}^{\Delta-1} (\hat h^{(0)}(\zeta)) =0.
\end{equation}
The problem of building the basis from (\ref{basiscond}) is complicated and in previous approaches  \cite{Alday:2015ewa, Alday:2016njk, Alday:2016jfr} this problem was solved as an expansion in $\zeta$. In the following we will show that for the case of integer $\Delta$ this problem can be solved systematically with the introduction of special functions.

\subsubsection{Harmonic polylogarithms and the basis $\hat h^{(n)}(\zeta)$}

Harmonic polylogarithms, labeled by some vector of zeroes and ones, can be defined in terms of nested integrals as follows \cite{Remiddi:1999ew}:
\begin{eqnarray}
H_0(\zeta)& =& \log (\zeta), \label{hpl1}\\
H_1(\zeta)& =& \log(1+\zeta),
\end{eqnarray}
together with
\begin{eqnarray}
H_{\vec{0}_n}(\zeta)& =&\frac{1}{n!} \log^n (\zeta), \\
H_{a \vec{\omega}}(\zeta)& =& \int_0^\zeta \frac{1}{a+x}  H_{\vec{\omega}}(x) dx\,.\label{hpl2}
\end{eqnarray}
where for the case at hand $a=0,1$. The first thing to notice is that $\hat {\cal C}$ has a very simple action on harmonic polylogarithms. For instance
\begin{eqnarray}
\hat {\cal C} H_{10 \vec{\omega}}(\zeta) &=& H_{0 \vec{\omega}}(\zeta)+H_{\vec{\omega}}(\zeta), \\
\hat {\cal C} H_{01 \vec{\omega}}(\zeta) &=& H_{1 \vec{\omega}}(\zeta)+H_{\vec{\omega}}(\zeta) \,.
\end{eqnarray}
We will find it convenient to introduce the following vector of interchanged zeroes and ones, of length $w$: $\vec{\rho}_w=(\cdots, 1,0, 1)$. Note that the last element is always $1$. The Casimir  $\hat {\cal C}$ has a very natural action on the weight-$w$ harmonic functions $H_{\vec{\rho}_w}(\zeta)$, where we define $H_{\vec{\rho}_0}(\zeta)=1$. We have
\begin{equation}
\label{ConH}
\hat {\cal C} H_{\vec{\rho}_w}(\zeta) =H_{\vec{\rho}_{w-1}}(\zeta) +H_{\vec{\rho}_{w-2}}(\zeta),
\end{equation}
together with 
\begin{equation}
\label{ConHborder}
\hat {\cal C} H_{\vec{\rho}_1}(\zeta) =1,~~~~~\hat {\cal C} H_{\vec{\rho}_0}(\zeta) =0\,.
\end{equation}
Furthermore, the functions $H_{\vec{\rho}_w}(\zeta)$ admit a series expansion around $\zeta=0$ with integer powers.  This leads to the following general solution to the necessary condition (\ref{neccond}):
\begin{equation}
\hat h^{(0)}_\Delta(\zeta)  = \alpha_0 H_{\vec{\rho}_0}(\zeta)+ \alpha_1 H_{\vec{\rho}_1}(\zeta) + \cdots + \alpha_{\Delta-2} H_{\vec{\rho}_{\Delta-2}}(\zeta)\,.
\end{equation}
For any integer $\Delta$ we have reduced the problem to finding a finite number of coefficients! In order to fix the coefficients we could plug this expression into the first relation of (\ref{basiscond}). Acting with $\hat {\cal C}^\Delta$  produces several divergent terms. Matching this divergence to $1/\zeta^\Delta$ we can fix all coefficients $\alpha_i$. There is, however, a more systematic way. In order to proceed, we would like to define the action of the inverse operator $\hat {\cal C}^{-1}$ on harmonic functions of the type considered here. From (\ref{ConH}) and  (\ref{ConHborder}) we obtain
\begin{equation}
\label{Cinv}
\hat {\cal C}^{-1}H_{\vec{\rho}_{w}}(\zeta) = \sum_{i=0}^w (-1)^i H_{\vec{\rho}_{w+1-i}}(\zeta)= H_{\vec{\rho}_{w+1}}(\zeta) - H_{\vec{\rho}_{w}}(\zeta) +H_{\vec{\rho}_{w-1}}(\zeta) + \cdots + (-1)^w H_{\vec{\rho}_{1}}(\zeta)  \,,
\end{equation}
where ambiguous terms which vanish under the action of $\hat {\cal C}$ have been chosen so as to have the right analytic properties on the right-hand side. Now, assume we have found  $\hat h^{(0)}_\Delta(\zeta)$ such that
\begin{equation}
\hat {\cal C}^{\Delta}(\hat h^{(0)}_\Delta(\zeta) \log^2 (\zeta))= \frac{1}{\zeta^\Delta} + \cdots
\end{equation}
We claim that
\begin{equation}
\label{recurrence}
\hat h^{(0)}_{\Delta+1}(\zeta) = \frac{1}{\Delta^2} \hat h^{(0)}_\Delta(\zeta) + \left(\frac{1}{\Delta}-1 \right) \hat {\cal C}^{-1}h^{(0)}_\Delta(\zeta)\,.
\end{equation}
Indeed, assuming this we can show
\begin{eqnarray}
\hat {\cal C}^{\Delta+1}(\hat h^{(0)}_{\Delta+1}(\zeta) \log^2 (\zeta))= \frac{1}{\Delta^2} \hat {\cal C} \frac{1}{\zeta^\Delta} + \left(\frac{1}{\Delta}-1 \right) \frac{1}{\zeta^\Delta}= \frac{1}{\zeta^{\Delta+1}}\,,
\end{eqnarray}%
while %
\e{}{\hat h^{(0)}_{2}(\zeta)=\frac{1}{2}H_{\vec{\rho}_0}(\zeta)=\frac{1}{2}~.}
This allows us to construct iteratively the solutions $\hat h^{(0)}_\Delta(\zeta) $ for any integer $\Delta$. The  inverse operator $\hat {\cal C}^{-1}$ also allows to compute
\begin{equation}
\label{nform}
\hat h^{(n)}_{\Delta}(\zeta) = \hat {\cal C}^{-n}\hat h^{(0)}_{\Delta}(\zeta) \,.
\end{equation}
The recurrence relations (\ref{recurrence}) and (\ref{nform}) together with the explicit expression for $\hat h^{(0)}_{2}(\zeta)$ provide the full solution to the problem for integer $\Delta$. In practice, it is convenient to have an expression for $\hat h^{(n)}_{\Delta}(\zeta)$ as a series expansion in $\zeta$. This can be obtained from $\hat h^{(0)}_{2}(\zeta)= \frac{1}{2}$ together with 
\begin{equation}
\hat {\cal C}^{-1} (\zeta^k) = \frac{\zeta^{k+1}}{(1+k)^2} ~_2F_1(1,1+k,2+k;-\zeta)\,.
\end{equation}
With this action plus the recurrence relations (\ref{recurrence}) and (\ref{nform}) we can generate series expansions to arbitrarily high orders. 

\subsection{$\gamma_{0,\ell}^{(2)}$ due to individual conformal blocks}

As discussed in Section \ref{crossingloops}, given a solution $\{\gamma_{n,\ell}^{(1)}, a_{n,\ell}^{(1)} \}$ to the crossing equations to order $1/N^2$, this generates a specific term at order $1/N^4$ proportional to $\log^2(u)$. In order to simplify our discussion we can further consider the term proportional to $\log v$ in a small $v$ expansion. More precisely, at leading order in the power expansion in $v$,
\es{logun4n}{
\left. {\cal G}^{(2)}(u,v) \right|_{\log^2 (u)} &= \sum_{n,\ell}u^{\Delta+n}  \frac{1}{8} a_{n,\ell}^{(0)} \left(\gamma_{n,\ell}^{(1)} \right)^2 g_{2\Delta+2n+\ell,\ell}(u,v)\\& \equiv u^\Delta f(u) \log v + O(v,\log v)\,.}
By crossing symmetry, there should be a corresponding term proportional to $\log u f(v) \log^2(v)$. For solutions with finite support in the spin at order $1/N^2$, this term can only by reproduced by $\gamma_{0,\ell}^{(2)}$, satisfying (\ref{gammaeq}).

We will split the problem in two parts. In this subsection, we will consider $f(u)$ produced by a single conformal block in the sum (\ref{logun4n}), and we will compute the corresponding $\gamma_{0,\ell}^{(2)}$ for such contribution. This part of the problem is universal. Then, for specific examples one can plug in the corresponding factor $a_{n,\ell}^{(0)} \left(\gamma_{n,\ell}^{(1)} \right)^2$ and perform the sum. This will be done in the next section. 

For a single conformal block we have, in two dimensions 
\begin{eqnarray}
f_{n,s}(u) &=& \left. u^n g_{2\Delta+2n+s,s}(u,v)\right|_{\log v} \\
&=& \frac{-u^n}{1+\delta_{\ell,0}} \left( \frac{\Gamma(2(n+s+\Delta))}{\Gamma^2(n+s+\Delta)} F_{n+\Delta}(u)+u^{s} \frac{\Gamma(2(n+\Delta))}{\Gamma^2(n+\Delta)} F_{n+s+\Delta}(u) \right)\,, \nonumber
\end{eqnarray}
while in four dimensions
\begin{eqnarray}
f_{n,s}(u) &=& \left. u^n g_{2\Delta+2n+s,s}(u,v)\right|_{\log v} \\
&=& \frac{u^n}{1-u}\left( -\frac{\Gamma(2(n+s+\Delta))}{\Gamma^2(n+s+\Delta)} F_{n+\Delta-1}(u)+u^{s+1} \frac{\Gamma(2(n+\Delta-1))}{\Gamma^2(n+\Delta-1)} F_{n+s+\Delta}(u) \right)\,, \nonumber
\end{eqnarray}
where again $F_\beta(x)=~_2F_1(\beta,\beta,2\beta;x)$. Note that the small $u$ behaviour is universal,
\begin{equation}
f_{n,s}(u)  = -\frac{\Gamma(2(n+s+\Delta))}{\Gamma^2(n+s+\Delta)} u^n + \cdots
\end{equation}
Having $f_{n,s}(u)$ we can apply the algebraic method developed in the previous section to  $f_{n,s}(v)$. The method is completely general, but the precise answers will depend on the number of dimensions, on $n$ and $s$, and on the parameter $\Delta$. It is convenient to write the result as follows:
\begin{equation}
f_{n,s}(v) \to \left. \gamma_{0,\ell}^{(2)} \right|_{(n,s)}= -\frac{c^{(0)}_{n,s}}{J^{2\Delta+2n}} \left(1+ \frac{c^{(1)}_{n,s}}{J^2}+\cdots \right)\equiv-\frac{c^{(0)}_{n,s}}{J^{2\Delta+2n}} \left. \hat \gamma_{0,\ell}^{(2)} \right|_{(n,s)}\,,
\end{equation}
where 
\begin{equation}
c^{(0)}_{n,s}=\frac{4 \Gamma(2n+2s+2\Delta)\Gamma^2(n+1)\Gamma^2(\Delta)}{\Gamma^2(n+s+\Delta)}\,.
\end{equation}
Recall that $J^2=(\Delta+\ell)(\Delta+\ell-1)$. $\hat \gamma_{0,\ell}^{(2)}$ has been defined in such a way that it starts at 1 in a large spin expansion. All the coefficients in the expansion of $ \left. \hat \gamma_{0,\ell}^{(2)} \right|_{(n,s)}$ can be computed by the method outlined above. For several examples we will be able to guess the function which is analytic in the half plane $\ell \geq 0$ and whose asymptotic expansion agrees with the expansion resulting from our method. Below we list those examples. 

\subsubsection*{Case 1: $d=2,\Delta=2$}
Let us start by analysing the effect of a single conformal block of spin zero, $s=0$. In this case we can resum the result, which has the following structure for arbitrary $\ell$:
\begin{equation}
\label{resummedfn0}
f_{n,0}(v) \to \left. \hat \gamma_{0,\ell}^{(2)} \right|_{(n,0)}= J^{2n+2} P_1^{2+2n}(\ell) + J^{2n+6} P_2^{2n}(\ell) \psi^{(2)}(\ell+1)\,,
\end{equation}
where $P_1^{2+2n}(\ell)$ and $P_2^{2n}(\ell)$ are polynomials in $\ell$ of degree $2+2n$ and $2n$, respectively. For this case $J^2=(\ell+1)(\ell+2)$.  Although we have not found a closed expression for them, they can be constructed for arbitrarily large $n$. 

Having found the re-summed expressions (\ref{resummedfn0}) for $ \left. \hat \gamma_{0,\ell}^{(2)} \right|_{(n,0)}$ we can actually extrapolate the results down to finite values of $\ell$. The polynomials $P_1^{2+2n}(\ell),P_2^{2n}(\ell)$ above have a very interesting structure when evaluated at integer values of $\ell$. For the present case $d=2, \Delta=2$ the results have the following structure
\es{}{
\frac{1}{J^{2n}} \left. \hat \gamma_{0,\ell}^{(2)} \right|_{(n,0)} = ~&\frac{4^n \Gamma \left(n+\frac{5}{2}\right)}{\sqrt{\pi } (n+1) \Gamma (n+2) \Gamma^2(n+3)}\\&\times\left(Q_1^{(2\ell+3)}(n)+(n+1)^3(n+2)^2Q_2^{(2\ell)}(n) \psi^{(2)}(n+1) \right)\,,}
where $Q_1^{(2\ell+3)}(n)$ and $Q_2^{(2\ell)}(n)$ are polynomials of degree $2\ell+3$ and $2\ell$, respectively. In other words, for integer values of $n$ $ \left. \hat \gamma_{0,\ell}^{(2)} \right|_{(n,0)}$ can be written in terms of polynomials in $\ell$, while for integer values of $\ell$ it can be written in terms of polynomials in $n$. 
The precise combination in parentheses can be seen to behave for large $n$ as 
\begin{eqnarray}
\label{largen2d}
Q_1^{(2\ell+3)}(n)+(n+1)^3(n+2)^2Q_2^{(2\ell)}(n) \psi^{(2)}(n+1) \sim \frac{1}{n^{2\ell+1}}\,.
\end{eqnarray}
This behaviour will play an important role in our discussion below.

The results above can be generalized to the effects of $f_{n,s}$, for generic $s$. In this case we obtain
\begin{equation}
f_{n,s}(v) \to \left. \hat \gamma_{0,\ell}^{(2)} \right|_{(n,s)}= J^{2n} P_1^{2n+2s+4}(\ell) + J^{2n+6} P_2^{2n+2s}(\ell) \psi^{(2)}(\ell+1)\,.
\end{equation}
As before, the results can also be extrapolated down to finite integer values of the spin. For instance, for $s=2$ we obtain
\es{}{\frac{1}{J^{2n}} \left. \hat \gamma_{0,\ell}^{(2)} \right|_{(n,2)} =~& \frac{4^n \Gamma \left(n+\frac{7}{2}\right)}{\sqrt{\pi } (n+1) \Gamma (n+4) \Gamma^2(n+5)}\\&\times\left( Q_1^{(2\ell+7)}(n)+\frac{\Gamma^3(n+4)}{\Gamma^3(n+1)}(4+n)^2Q_2^{(2\ell-2)}(n) \psi^{(2)}(n+1) \right)\,.}
For $\ell=0$, $Q_2(n)=0$ and $Q_1(n)$ is a polynomial of cubic order. Furthermore, one can check
\begin{eqnarray}
\label{largend2s2}
Q_1^{(2\ell+7)}(n)+\frac{\Gamma^3(n+4)}{\Gamma^3(n+1)}(4+n)^2Q_2^{(2\ell-2)}(n) \psi^{(2)}(n+1) \sim \frac{1}{n^{2\ell-3}}\,.
\end{eqnarray}

\subsubsection*{Case 2: $d=4,\Delta=2$}

In this case, and for $s=0$, the results have the following structure:
\begin{equation}
f_{n,0}(v) \to \left. \hat \gamma_{0,\ell}^{(2)} \right|_{(n,0)} = J^{2n-2} P_1^{2n+4}(\ell) + J^{2n+4} P_2^{2n}(\ell) \psi^{(2)}(\ell+1)\,,
\end{equation}
where $P_1^{2n+4}(\ell)$ and $P_2^{2n}(\ell)$ are polynomials in $\ell$ of degree $2n+4$ and $2n$, respectively, not necessarily equal to the polynomials above. The explicit results for the first few values of $n$ are included in the appendix \ref{appb}. Again, they satisfy the same constraints as for the $d=2$ case. As before, having computed the resummed expressions, we can extrapolate them down to finite values of the spin. For the present case $d=4, \Delta=2$ the results have the following structure
\begin{eqnarray}
\frac{1}{J^{2n}} \left. \hat \gamma_{0,\ell}^{(2)} \right|_{(n,0)} = \frac{4^n \Gamma \left(n+\frac{3}{2}\right)}{\sqrt{\pi }\Gamma^3(n+2)}\left( Q_1^{(2\ell+2)}(n)+(n+1)^4Q_2^{(2\ell)}(n) \psi^{(2)}(n+1) \right)\,,
\end{eqnarray}
where $Q_1^{(2\ell+2)}(n)$ and $Q_2^{(2\ell)}(n)$ are polynomials of degree $2\ell+2$ and $2\ell$, respectively. Again, the specific combination appearing above satisfies
\begin{equation}
\label{largen4d}
Q_1^{(2\ell+2)}(n)+(n+1)^4Q_2^{(2\ell)}(n) \psi^{(2)}(n+1) \sim \frac{1}{n^{2\ell+2}}
\end{equation}
for large $n$, which imposes strong constraints. Again, the results above can be generalized to the exchange of a more general conformal block, with $s \neq 0$, in the dual channel. For $\hat \gamma_{0,\ell}^{(2)}$ we obtain
\begin{equation}
f_{n,s}(v) \to \hat \gamma_{0,\ell}^{(2)} = J^{2n-2} P_1^{2n+2s+4}(\ell) + J^{2n+2} P_2^{2n+2s+2}(\ell) \psi^{(2)}(\ell+1)\,.
\end{equation}
For $s=2$ we have found the extrapolation of the above results to finite values of the spin, for general $n$. The results have the structure: 
\begin{eqnarray}
\frac{1}{J^{2n}} \left. \hat \gamma_{0,\ell}^{(2)} \right|_{(n,2)} = \frac{4^n \Gamma \left(n+\frac{3}{2}\right)}{\sqrt{\pi }\Gamma^3(n+4)}\left( Q_1^{(2\ell+8)}(n)+\frac{\Gamma^3(n+4)}{\Gamma^3(n+1)}(2+n) Q_2^{(2\ell)}(n) \psi^{(2)}(n+1) \right)\,.
\end{eqnarray}
 Furthermore, one can check the following behaviour at large $n$:
 \begin{eqnarray}
Q_1^{(2\ell+8)}(n)+\frac{\Gamma^3(n+4)}{\Gamma^3(n+1)}(2+n) Q_2^{(2\ell)}(n) \psi^{(2)}(n+1) \sim \frac{1}{n^{2\ell-4}}\,.
\end{eqnarray}
 
\ssec{Summary}
We are now ready to assemble all the ingredients together and compute $\gamma_{0,\ell}^{(2)}$ for specific examples, which we will do in the next section. First, let's recap. As discussed in Section \ref{crossingloops}, given a solution $\{\gamma_{n,\ell}^{(1)}, a_{n,\ell}^{(1)} \}$ to the crossing equations to order $1/N^2$, this generates a specific term at order $1/N^4$ proportional to $\log^2(u)$. In order to simplify our discussion we further considered the term proportional to $\log v$ in a small $v$ expansion. More precisely, at leading order in the power expansion in $v$,
\begin{equation}
\label{logun4}
\left. {\cal G}^{(2)}(u,v) \right|_{\log^2 (u)} = \sum_{n,\ell}u^{\Delta+n}  \frac{1}{8} a_{n,\ell}^{(0)} \left(\gamma_{n,\ell}^{(1)} \right)^2 g_{2\Delta+2n+\ell,\ell}(u,v) \equiv u^\Delta f(u) \log v + O(v,\log v)\,.
\end{equation}
By crossing symmetry, there should be a corresponding term proportional to $\log u f(v) \log^2(v)$. For solutions with finite support in the spin at order $1/N^2$, this term can only by reproduced by $\gamma_{0,\ell}^{(2)}$, which should be such that 
\begin{equation}
\sum_\ell a_{0,\ell}^{(0)} \gamma_{0,\ell}^{(2)} g_{2\Delta+\ell,\ell}^\coll(v) = 2 f(v) \log^2 (v)+ \cdots
\end{equation}
$f(v)$ receives contributions from all double-trace operators $\OO_{n,s}$ whose anomalous dimension is different from zero at order $1/N^2$. We call such contribution $f_{n,s}(v)$, and we denote their contributions to the anomalous dimension $\gamma_{0,\ell}^{(2)} \big|_{(n,s)}$. Given the anomalous dimensions $\gamma_{n,s}^{(1)}$ for double trace operators at order $1/N^2$, the total contribution to $\gamma_{0,\ell}^{(2)}$ is given by
\begin{equation}
\label{finalgamma}
\gamma_{0,\ell}^{(2)}= \frac{1}{8} \sum_{n,s} a_{n,s}^{(0)} \left(\gamma_{n,s}^{(1)} \right)^2 \left. \gamma_{0,\ell}^{(2)} \right|_{(n,s)}\,,
\end{equation}
where the computation of $\gamma_{0,\ell}^{(2)} \big|_{(n,s)}$ was described in previous subsections.

We make one final note before moving on, which is the issue of convergence when summing over $n$. Let us for simplicity consider the case $s=0$. Given the behaviour (\ref{largen2d}) and (\ref{largen4d}) and the explicit form of the OPE coefficients at tree level, we obtain the large $n$ behaviour
\begin{eqnarray}
 a_{n,0}^{(0)}\left. \gamma_{0,\ell}^{(2)} \right|_{(n,0)} \sim \frac{1}{n^{2\ell}},~~~~\text{for $d=2$}\,,\\
 a_{n,0}^{(0)}\left. \gamma_{0,\ell}^{(2)} \right|_{(n,0)} \sim \frac{1}{n^{2\ell+2}},~~~~\text{for $d=4$}\,.
\end{eqnarray}
For a fixed value of $\ell$, the convergence will depend on the behaviour of $\gamma_{n,s}^{(1)}$ for large values of $n$.

\sec{Explicit examples}\label{s5}
We first focus on the poster child for an effective theory in AdS dual to a generalized free field in CFT, namely, the $\phi^4$ theory. We then move on to an example for which no one-loop data has been computed in AdS: $\phi^3+\phi^4$ theory. 

\begin{figure}[t!]
   \begin{center}
 \includegraphics[width = 0.45\textwidth]{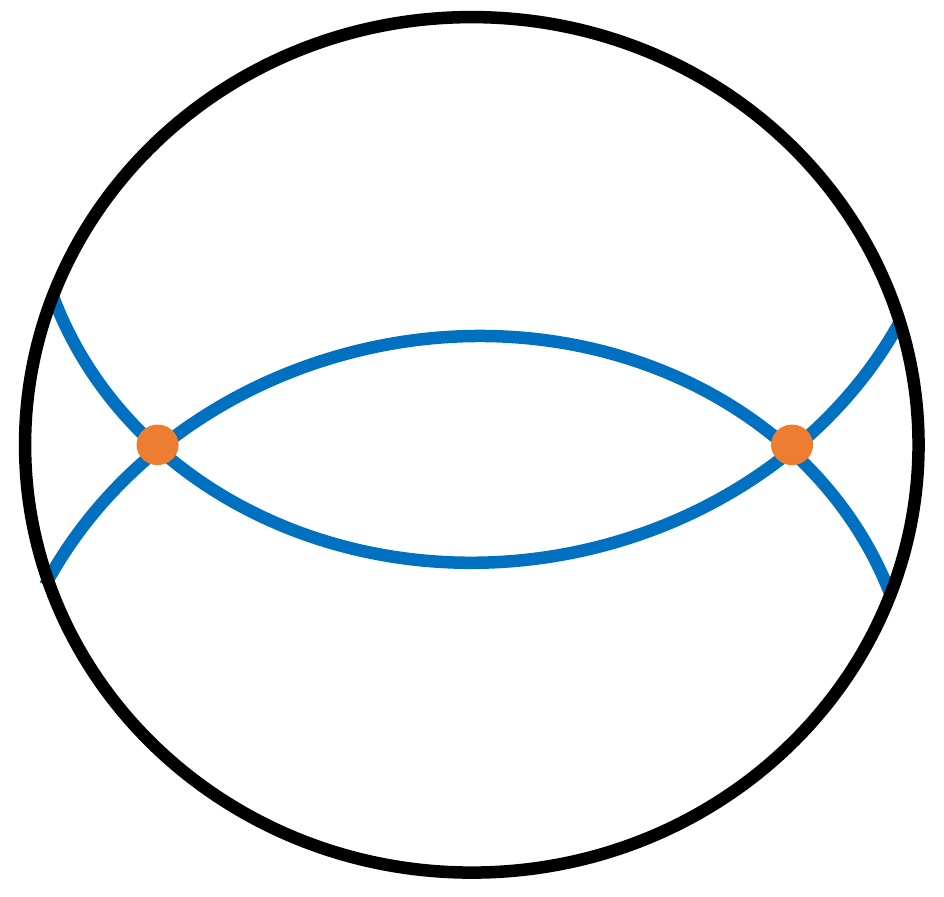}
 \caption{The one-loop bubble diagram in AdS $\phi^4$ theory. The corresponding Mellin amplitude is given in \eqr{m1loopads}.}
 \label{bubble}
 \end{center}
 \end{figure}

\ssec{$\phi^4$ in AdS}
This is the simplest solution of crossing at $O(1/N^2)$, as constructed in \cite{Heemskerk:2009pn}, because it has support only for spin zero. In AdS, there is just one non-trivial diagram (in each channel), namely, the scalar bubble diagram shown in Figure \ref{bubble}. Our goal here is concrete: we seek to reconstruct the AdS amplitude from crossing symmetry by computing anomalous dimensions, and to detect the UV divergence in $d\geq 3$, i.e. AdS$_{D\geq 4}$. Upon doing so, we will show perfect agreement with previous calculations.  

\sssec{Expectations from the bulk}
Let us set our expectations, discussed in the Introduction and Appendix \ref{UVdiv}, regarding UV divergences. The $\l\phi^4$ theory will diverge in AdS$_{D\geq 4}$, just like its flat space counterpart. In $4\leq D \leq 7$, the divergence is cured by a $\phi^4$ counterterm; local counterterms with more derivatives, such as $(\p\phi)^4$, are not required in this range of $D$ (but will be required at higher $D$). Because a $\phi^4$ coupling only generates anomalous dimensions for $\ell=0$, this implies that in a CFT calculation at $O(1/N^4)$, there should be a sharp signature of the divergence: only the $\ell=0$ OPE data should diverge, but all $\ell>0$ data should be finite. 

Recall that in general when we solve the crossing equation, we have a full family of solutions that differ by local four-point bulk couplings. So if we compute some one-loop four-point function by a direct evaluation in AdS (with some counter-terms that make it finite), and compare it to some solution that we choose for the crossing equation, there is a priori no need for any specific $\gamma^{(2)}_{n,\ell}$ to match: only the $1/\ell$ (or $1/J^2$) expansion must match at all orders. However, in the CFT methods employed herein, there is a natural way to resum the $1/\ell$ expansion of $\gamma^{(2)}_{n,\ell}$ to an analytic function of $\ell$. Since these methods yield analytic solutions which do not treat $\ell=0$ differently than any other value, it is natural to match them to a bulk computation without any explicit counter-terms. One argument for this is that for $d<3$ the bulk diagram converges, so we expect the results, both from the bulk and from crossing, to be analytic in $\ell$ for all values of $\ell$. However, the bulk computation, and also the crossing computation when properly defined, are analytic in $d$, so if they agree (without any counter-terms) for some range of values of $d$, they will agree for all values. We will see that indeed this naive expectation is realized, and we will find a divergence at $d=4$ precisely for $\ell=0$.\footnote{When we compute with some finite cutoff in the bulk, we also expect to have in the effective theory higher derivative terms suppressed by this cutoff scale, that should give additional contributions at all values of $\ell$, suppressed by the cutoff scale. In particular, this would be the case if there are additional massive particles that we are neglecting. However, we see that when we ignore these effects at the leading order, and naively take the cutoff to infinity, we can ignore the details of the cutoff also at the one-loop order, and obtain precise results for any quantities that are not UV-divergent.}

\sssec{Solution from crossing}
The anomalous dimensions at order $1/N^2$ are given by
\es{gammaphi4}{d=2:&\quad\quad \gamma_{n,0}^{(1)} = \frac{2\Delta-1}{2\Delta+2n-1} \alpha \\
d=4:&\quad\quad \gamma_{n,0}^{(1)} = \frac{(2 \Delta -1) (n+1) (2 \Delta +n-3) (\Delta +n-1)}{(\Delta -1) (2 \Delta +2 n-3) (2 \Delta +2 n-1)} \alpha~,}
where we have set 
\e{}{\gamma_{0,0}^{(1)}  \equiv \alpha\,.}
For simplicity, we take $\D=2$ in what follows.

In two dimensions $\gamma_{n,0}^{(1)}  \sim \frac{1}{n}$ and the sum \eqref{finalgamma} over $n$ is convergent for $\ell=0,2,\cdots$. Performing explicitly the sum (\ref{finalgamma}) for the first few cases we obtain
\begin{eqnarray}
\gamma_{0,0}^{(2)} &=& \frac{15-4\pi^2}{10} \alpha^2\,,\label{d2l0} \\
\gamma_{0,2}^{(2)} &=& \frac{455-48\pi^2}{420} \alpha^2\,,\label{d2l2}\\
\gamma_{0,4}^{(2)} &=& \frac{5863-600\pi^2}{8580} \alpha^2\,.\label{d2l4}
\end{eqnarray}

In four dimensions the situation is a bit different. Since $\gamma_{n,0}^{(1)}  \sim n$ (see Appendix \ref{UVdiv}), the sums above are divergent for $\ell=0$ and convergent for $\ell=2,4,\cdots$, as expected. In this case we obtain\footnote{These and similar sums can be performed with the help of the following two relations $$\sum_{n=0}^\infty \psi^{(2)}(n+1) e^{n \epsilon}=\frac{2(Li_3(e^\epsilon)-\zeta_3)}{1-e^{\epsilon}},~~~~~\sum_{n=0}^\infty \frac{\psi^{(2)}(n+1)}{(2n+1)(2n+3)}=8-2\zeta_2-8 \log 2,$$ together with derivatives of the first.}
\begin{eqnarray}
\gamma_{0,0}^{(2)} &\to& \text{divergent}\,, \\
\gamma_{0,2}^{(2)} &=& \frac{2(174\pi^2-1925)}{3465} \alpha^2\,,\label{d4l2}\\
\gamma_{0,4}^{(2)} &=& \frac{150600\pi^2 -1520519 }{2252250} \alpha^2\,. \label{d4l4}
\end{eqnarray}

The above results were obtained by resummation of the large spin expansion. As explained in Section \ref{s3cross}, we can also use the large spin expansion of $\g^{\2}_{0,\ell}$ together with \eqr{g2phi4} to reconstruct $\mloop$, pole-by-pole. For $d=4$ we obtain
\begin{equation}\label{d4large}
\gamma_{0,\ell}^{(2)} = -\frac{12}{J^4}\left(1+\frac{18}{5} \frac{1}{J^2}+\frac{96}{7} \frac{1}{J^4}+\frac{360}{7} \frac{1}{J^6} +\frac{74304}{385} \frac{1}{J^8}+\frac{724320}{1001}\frac{1}{J^{10}}+\cdots \right)\alpha^2 \,.
\end{equation}
Symmetrizing over channels and extending the $1/J$ expansion \eqr{3f2asy} and \eqr{3f2asy2} to higher orders, we infer the following result for the amplitude:
\e{mloopform}{\mloop(s,t) = \sum_{m=0}^\i {R_m\over t-(4+2m)} + \text{(crossed)}\,,}
with 
\e{d4res}{R_m =- {9(3m+4)4^m(m+1)!^2\over (2m+3)!}\a^2\,.}
This corresponds to an AdS amplitude with vanishing regular part, $f_{\reg}=0$, since we have only reconstructed its polar part. As explained above, this is closely related to the infinity we found for $\g^{\2}_{0,0}$ in $d=4$. For $d=2$, the large spin expansion is
\begin{equation}\label{d2large}
\gamma_{0,\ell}^{(2)} = -\frac{6}{J^4}\left(1+\frac{4}{5} \frac{1}{J^2}+\frac{4}{7} \frac{1}{J^4}+\frac{16}{35} \frac{1}{J^6} +\frac{16}{55} \frac{1}{J^8}+\frac{1856}{5005}\frac{1}{J^{10}}+\cdots \right)\alpha^2 ~,
\end{equation}
and the resulting $\mloop$ is again of the form \eqr{mloopform}, with
\e{d2res}{R_m = -{9(3m+4)4^m(m+1)!^2\over 2(m+1)(2m+3)!}\a^2\,.}

\sssec{Comparison to AdS results}
In \cite{Penedones:2010ue} and then in \cite{Fitzpatrick:2011dm}, the bubble diagram was computed directly in AdS $\phi^4$ theory by utilizing a trick of harmonic analysis in AdS: one can replace a product of two bulk-to-bulk propagators that start and end at the same point by an infinite sum of single propagators, thus reducing the bubble diagram to an infinite sum of computable tree-level exchange diagrams. The result of \cite{Fitzpatrick:2011dm}, in the $t$-channel, say, was
\e{m1loopads}{\mloop(s,t)= \sum_{m=0}^\i {\hat R_m\over t-(2\D+2m)}\,,}
with
\e{}{\hat R_m = \sum_{p=0}^mN_{\D}(p) r_{m-p}\,,}
where
\es{}{N_{\D}(p) &= \frac{\pi ^{-\frac{d}{2}} \left(\frac{d}{2}\right)_p (-d+p+2 \Delta +1)_p (2 p+2 \Delta )_{1-\frac{d}{2}}}{2 p! (p+\Delta )^2_{1-\frac{d}{2}} \left(-\frac{d}{2}+p+2 \Delta \right)_p}\,,\\
r_{m-p} &= -\frac{\pi ^{-\frac{3 d}{2}} (p+1)^2_{m-p}\, \Gamma^2 \left(-\frac{d}{2}+p+2 \Delta \right)}{64 (m-p)! \,\Gamma^4 \left(-\frac{d}{2}+\Delta +1\right) \Gamma \left(-\frac{d}{2}+m+p+2 \Delta +1\right)}\,.}
Their conventions use a specific value for $\a^2$. We can extract that value by matching the large spin asymptotics of $\g^{\2}_{0,\ell}$, as computed from the above, to \eqr{d4large} and \eqr{d2large}. From \eqr{g2large2}, the large spin asymptotics is
\e{}{\g^{\2}_{0,\ell\gg 1} \sim -2\Gamma^4(\D)\hat R_0 \ell^{-2\D}\,.}
For $d=4, \D=2$, upon matching to \eqr{d4large}, we require $2\hat R_0 = 12\a^2$, which defines the $\a^2$ normalization of \cite{Fitzpatrick:2011ia}.\footnote{\label{foot7}Explicitly, $\a^2 = 1/4608\pi^8$ in $d=4$, and $\a^2 = 1/576\pi^4$ in $d=2$.}
We can then write the residues $\hat R_m$ as
\e{d4resf}{d=4~, ~ \D=2:\quad \hat R_m = -{9(3m+4)4^m(m+1)!^2\over (2m+3)!}\a^2~.}
This matches \eqr{d4res}. In $d=2, \D=2$, upon matching to \eqr{d2large}, we require $2\hat R_0 = 6\a^2$, which yields
\e{}{d=2~, ~ \D=2:\quad \hat R_m = -{9(3m+4)4^m(m+1)!^2\over 2(m+1)(2m+3)!}\a^2~.}%
This matches \eqr{d2res}. This is a substantial check on the match between CFT and AdS: we have successfully reconstructed the $\phi^4$ one-loop amplitude from the conformal bootstrap.

The expected UV divergence structure is apparent in the above: at large $m$, one has
\e{uvdiv}{\hat R_m \sim m^{d-3\over 2}\,.}%
This leads to a divergence in the sum over $m$ for $d\geq 3$ -- that is, in AdS$_{D\geq 4}$ -- with a logarithmic divergence at the critical dimension $d_c=3$. We note for later that the $d=2$ amplitude can be resummed to yield the $t$-channel amplitude
\e{mloopd2}{M_{\rm 1-loop}(s,t) = -3\left(\frac{\, _3F_2\left(1,1,2-\frac{t}{2};\frac{5}{2},3-\frac{t}{2};1\right)}{t-4}+\frac{3}{10}\frac{ \, _3F_2\left(2,2,3-\frac{t}{2};\frac{7}{2},4-\frac{t}{2};1\right)}{t-6}\right)\a^2~.}

It is not obvious that the low-spin $\g^{\2}_{0,\ell}$ as computed from the Mellin amplitudes above will match those from the crossing problem. For this reason, we would like to analytically compute $\g^{\2}_{0,\ell}$ for $\ell=2,4$ directly from \eqr{m1loopads}. This has never been done. Doing so requires new techniques that should be useful more generally for extracting anomalous dimensions from Mellin amplitudes with an infinite series of poles in a given channel. We devote Section \ref{s6} to this endeavor. The end result is a perfect match for $\ell=2,4$ in both $d=2$ and $d=4$.

\sssec{Relation to lightcone bootstrap}

Note that in both $d=2$ and $d=4$, the anomalous dimensions are negative and monotonically increasing with $\ell$: 
\e{}{\g^{\2}_{0,\ell}<0~, \quad {\p\g^{\2}_{0,\ell}\over \p \ell}>0\,.} 
We have checked this behavior to higher $\ell$ as well. These properties must in fact hold for all $\ell$ and all (unitary) $\D$, as can be explained by resorting to Nachtmann's theorem and the lightcone bootstrap. The basic point is that, because $\g^{\1}_{0,\ell>0}=0$, these one-loop anomalous dimensions are actually the leading corrections to the mean field theory result. The $\O\times\O$ OPE is reflection positive and contains only even spin operators, which implies monotonicity via the arguments in \cite{Komargodski:2012ek,Fitzpatrick:2012yx, Li:2015rfa}; moreover, the negativity follows from the large spin asymptotics given in \eqr{phi4large}. 
\sssec{More general contact interactions}
We could also consider more general solutions, where $\gamma_{n,s}^{(1)}$ is different from zero also for $s \neq 0$. This corresponds to $(\p\phi)^4$-type theories, etc. Let us analyse the issue of divergences in this case. For instance, for $s=2$ all the explicit results we have obtained (too cumbersome to be included here) are consistent with
\begin{eqnarray}
 a_{n,2}^{(0)}\left. \gamma_{0,\ell}^{(2)} \right|_{(n,2)} \sim \frac{1}{n^{2\ell+1}},~~~~\text{for $d=2$,}\\
 a_{n,2}^{(0)}\left. \gamma_{0,\ell}^{(2)} \right|_{(n,2)} \sim \frac{1}{n^{2\ell+2}},~~~~\text{for $d=4$.}
\end{eqnarray}
On the other hand, $\gamma_{n,s}^{(1)}$ has generally an enhanced behaviour, with respect to the $\phi^4$ solution studied above. For instance, an irrelevant interaction such as $(\partial \phi)^4$ leads to a behaviour $\gamma_{n,s}^{(1)} \sim n^{d+1}$, see \cite{Fitzpatrick:2012cg}, as expected from the analysis of Appendix \ref{UVdiv}. For $d=2$ this implies that the resulting $\gamma_{0,\ell}^{(2)}$ will be convergent only for $\ell>2$. For $d=4$, the result will be convergent only for $\ell>4$. 

\subsection{$\phi^3 + \phi^4$ in AdS}
\begin{figure}[t!]
   \begin{center}
 \includegraphics[width = 0.45\textwidth]{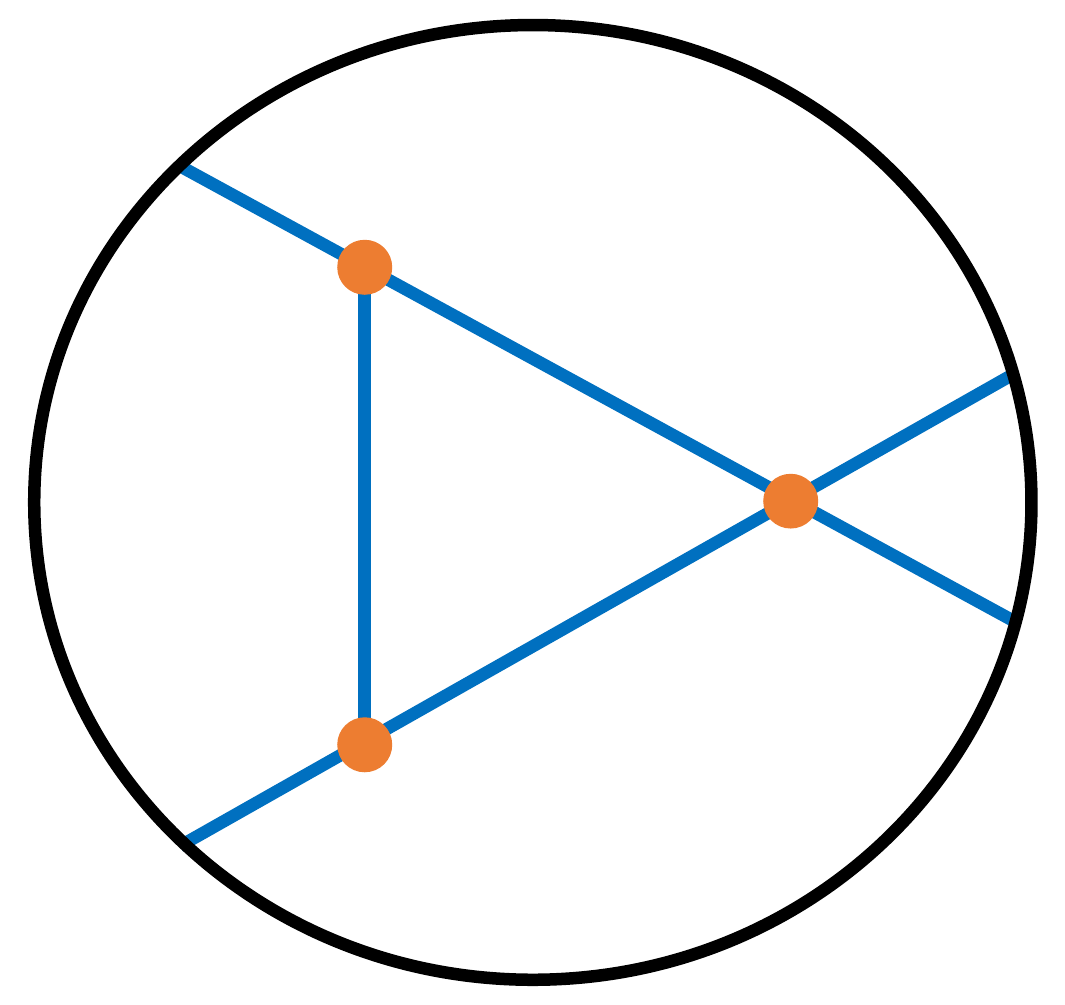}
 \caption{The one-loop triangle diagram in AdS $\phi^3+\phi^4$ theory. Together with the $\phi^4$ renormalization of $\phi^3$ tree-level diagrams, the corresponding Mellin amplitude for a $m^2=-4$ scalar in AdS$_5$ is given in \eqr{mlooptri2}. The amplitude for a massless scalar in AdS$_3$ is given in \eqr{mlooptrid2}.}
 \label{triangle}
 \end{center}
 \end{figure}

We now consider the following (Euclidean) AdS$_{d+1}$ effective theory:
\e{}{{\cal L}_{\rm bulk} = {1\over 2} (\p\phi)^2  +{1\over 2}m^2\phi^2 + {\mu_3\over 3!} \phi^3+{\mu_4\over 4!} \phi^4~.}
On the crossing side, we now consider in more detail the solution $\gamma^{(1),\phi^3}_{n,\ell}$ discussed in Section \ref{crossingloops}, together with a truncated $\phi^4$ solution $\gamma^{(1),\phi^4}_{n,\ell}$ with support only on operators with $\ell=0$. We will compute the crossed term contribution, proportional to $\mu_3^2 \mu_4$, to $\gamma_{0,\ell}^{(2)}$, namely
\begin{equation}
\label{crossedgamma}
\gamma_{0,\ell}^{(2)}= \frac{1}{4} \sum_{n} a_{n,0}^{(0)} \gamma^{(1),\phi^3}_{n,0}\gamma^{(1),\phi^4}_{n,0} \left. \gamma_{0,\ell}^{(2)} \right|_{(n,0)}\,.
\end{equation}
This only receives spin-0 contributions. Holographically, this computes the sum over channels of all diagrams in AdS$_5$ involving two cubic and one quartic vertex. The non-trivial diagram is the four-point triangle diagram in AdS$_5$, as shown in Figure \ref{triangle}.\footnote{One virtue of our crossing symmetry approach is that it yields the complete AdS amplitude without having to split it into diagrams. In the bulk dual of our computation, in addition to the triangle, there are also diagrams at order $\mu_3^2\mu_4$ that renormalize the $\phi^3$ tree-level exchange: namely, the bubble vertex correction and mass renormalization. Note that the Mellin space analytic structure of these renormalizations is not immediately obvious: vertices are integrated over all of AdS, so there is no obvious notion of one-particle-irreducibility. In particular, even these diagrams may have double-trace poles. The full sum over double-trace poles is the meaning of $M_{\rm 1-loop}$ computed below. Fortunately, the renormalization diagrams may also be computed directly in the bulk using existing technology, and subtracted to give the pure triangle amplitude. For instance, the bubble vertex correction (i.e. a tree-level exchange diagram where one vertex is replaced by a bubble with one cubic and one quartic vertex) may be computed in two steps: first use the Kallen-Lehmann rule in equation (90) of \cite{Fitzpatrick:2011hu} to write the bubble as a sum of single propagators; then write the resulting product of two bulk-to-bulk AdS propagators integrated over a common point as a sum of individual propagators (as in Section 4.2 of \cite{Hijano:2015zsa}). 

The above topic is addressed in \cite{mps} which includes a more detailed development of the methods herein and a proper treatment of the vertex and bulk-to-bulk mass renormalization diagrams, which do indeed contribute to $M_{\rm 1-loop}$. 

We thank Ellis Ye Yuan for stimulating discussions related to the above.}

For definiteness, we first take $d=4$ and $\Delta=2$, hence $m^2 =\D(\D-d)= -4$ in AdS units. The first-order data needed on the right-hand side is
\begin{equation}\label{phi3g1n0}
\gamma^{(1),\phi^3}_{n,0} = \mu_3^2 \frac{2(2-7(1+n)^2)}{(1+n)(3+4n(2+n))},~~~~~~\gamma^{(1),\phi^4}_{n,0} =\mu_4 \frac{3(n+1)^3}{(1+2n)(3+2n)}\,, 
\end{equation}
where $\gamma^{(1),\phi^3}_{n,0}$ is computed in Appendix \ref{appd}. We note that $\gamma^{(1),\phi^3}_{n,0}\gamma^{(1),\phi^4}_{n,0} \sim 1$ for large $n$. This leads to a convergent contribution  for (\ref{crossedgamma}) even for the case $\ell=0$, as expected from the bulk. For the first few spins we obtain
\begin{eqnarray}\label{phi3ad}
\gamma_{0,0}^{(2)} &=&  \frac{16}{5} \pi^2 \mu_3^2 \mu_4\,, \\
\gamma_{0,2}^{(2)} &=& \frac{16}{945} \left(39 \pi ^2-350\right) \mu_3^2 \mu_4\,,\\
\gamma_{0,4}^{(2)} &=& 8\left(\frac{62 \pi ^2}{1287}-\frac{1253}{2700} \right)\mu_3^2 \mu_4\,.
\end{eqnarray}
In addition, the full theory will have a term proportional to $\mu_4^2$ from the bubble diagram of Figure \ref{bubble}, which is exactly as before, plus a term proportional to $\mu_3^4$ from the box diagram of Figure \ref{loopexp}, which is harder to compute. 

The same analysis of \eqref{crossedgamma} can be done also for $d=2$. We again take $\D=2$, and hence $m^2=0$ in AdS units. Now we use
\begin{equation}\label{phi3cftd2}
\gamma^{(1),\phi^3}_{n,0} =- \mu_3^2 \frac{2 (4 n+5)}{(n+1) (n+2) (2 n+3)},~~~~~~\gamma^{(1),\phi^4}_{n,0} =\mu_4 \frac{3}{(2n+3)}\,.
\end{equation}
Analogously to the previous case, we compute the value of $\gamma^{(2)}$ for some value of the spin 
\begin{eqnarray}\label{phi3ad}
\gamma_{0,0}^{(2)} &=& \frac{4}{3} \left(\pi ^2-6\right) \mu_3^2 \mu_4\,, \\
\gamma_{0,2}^{(2)} &=&\frac{1}{14} \left(8 \pi ^2-77\right) \mu_3^2 \mu_4\,,\\
\gamma_{0,4}^{(2)} &=&\left(\frac{4 \pi ^2}{11}-\frac{107}{30}\right)\mu_3^2 \mu_4\,.
\end{eqnarray}

 As for the $\phi^4$ case, we may use the $1/J$ expansion of $\g^{\2}_{0,\ell}$ to reconstruct the bulk amplitude. The $1/J$ expansion used to derive the above results for $d=4$ is
\begin{equation}\label{largeJp34}
\gamma_{0,\ell}^{(2)} = \frac{80}{J^4}\left(1+\frac{18}{25} \frac{1}{J^2}+\frac{96}{175} \frac{1}{J^4}+\frac{72}{175}\frac{1}{J^6} +\frac{576}{1925} \frac{1}{J^8}+\frac{6624}{25025}\frac{1}{J^{10}}+\cdots \right)\mu_3^2 \mu_4 \,.
\end{equation}
Analogously to the $\phi^4$ case, we find the following result:
\e{mloopformp3}{\mloop(s,t) = \sum_{m=0}^\i {R_m\over t-(4+2m)} + \text{(crossed)}\,,}
with 
\begin{equation}
R_m=\frac{3(10+7m)\sqrt{\pi}\Gamma(1+m)}{\Gamma(\frac{5}{2}+m)}\mu_3^2 \mu_4\,.
\end{equation}
The amplitude can be resummed: in the $t$-channel, say,
\e{mlooptri2}{\mloop(s,t) = \left(\frac{40 \, _3F_2\left(1,1,2-\frac{t}{2};\frac{5}{2},3-\frac{t}{2};1\right)}{t-4}+\frac{56}{5}\frac{ \, _3F_2\left(2,2,3-\frac{t}{2};\frac{7}{2},4-\frac{t}{2};1\right)}{t-6}\right)\mu_3^2\mu_4}
as quoted in the introduction. Upon subtracting the renormalization (proportional to $\mu_3^2\mu_4$) of the tree-level exchange, this gives a prediction for the triangle Witten diagram for a $m^2=-4$ scalar in AdS$_5$. Note the striking similarity to the $\phi^4$ bubble diagram for $d=\D=2$ in \eqr{mloopd2}, which is completely unobvious from the spacetime perspective.

We can perform the same analysis for $d=2$ and $\Delta=2$. The only difference with respect to the previous case is $\gamma^{(1),\phi^3}_{n,0}$, which is computed in Appendix \ref{appd}.  The $1/J$ expansion is
\begin{equation}
\gamma_{0,\ell}^{(2)} = \frac{10}{J^4}\left(1+\frac{2}{25} \frac{1}{J^2}-\frac{8}{175} \frac{1}{J^4}+\frac{8}{175}\frac{1}{J^6} -\frac{32}{385} \frac{1}{J^8}+\frac{6112}{25025}\frac{1}{J^{10}}+\cdots \right)\mu_3^2 \mu_4 \,.
\end{equation}
In this case, we find \eqr{mloopformp3} with
\begin{equation}
R_m=\left(\frac{6}{1+m}-\frac{3 \sqrt{\pi} \Gamma(1+m)}{4 \Gamma(\frac{5}{2}+m)}\right)\mu_3^2 \mu_4\,.
\end{equation}
We can again resum the amplitude and obtain, in the $t$-channel,
\begin{equation}\label{mlooptrid2}
\mloop(s,t) =\left(\frac{6 H(1-\frac{t}{2})}{t-2}-\frac{\, _3F_2\left(1,1,2-\frac{t}{2};\frac{5}{2},3-\frac{t}{2};1\right)}{t-4}\right)\mu_3^2 \mu_4\,,
\end{equation}
where $H(x)$ denotes the harmonic number of argument $x$, defined for $x\notin\mathbb{Z}$ via the relation to the digamma function, $H(x) = \psi(x+1)+\gamma$.  Upon subtracting the renormalization of the tree-level exchange, this gives a prediction for the triangle Witten diagram for a massless scalar in AdS$_3$.

\sec{Computing anomalous dimensions from Mellin amplitudes}\label{s6}
In this section, we develop techniques for analytically computing double-trace anomalous dimensions from Mellin amplitudes. In particular, we focus on cases where the amplitude has an infinite series of poles. This necessarily occurs at one-loop as explained in this work, but also occurs in the tree-level exchange diagram of $\phi^3$ for generic $\D$, or the tree-level exchange of a dimension $\D'$ scalar between external dimension $\D$ scalars, where $\D'-2\D\notin 2\mathbb{Z}$. To our knowledge, the only treatments that have appeared in previous literature deal with finite sums of poles. As an application, we derive the one-loop $\phi^4$ anomalous dimensions for $\ell=2,4$, described in the previous section.

\ssec{General problem}
Consider an exchange amplitude between identical external scalars of dimension $\D$, of the form
\e{}{M(s,t) = \sum_{m=0}^\i {R_m(t)\over s-\D'-2m} + \text{(crossed)}}
for some residues $R_m$ and some internal dimension $\D'$. If this represents a tree-level exchange of a dimension $\D'$ scalar primary, say, the residues are $R_m(t)\propto \Q_{m,0}(t;\D)$. Define the integral
\es{mellindelta}{I_{\ell}(\D,\d)\equiv {1\over 2\pi i}\int_{-i\i}^{i\i} ds\, \left({1\over s-\d}\right)\G^2\left({s\over 2}\right)\G^2\left({2\D-s\over 2}\right){}_3F_2(-\ell,\ell+2\D-1,{s\over 2};\D,\D;1)\,.}
The double-trace anomalous dimension $\g^{\1}_{0,\ell>0}$ receives contributions from the two crossed channels: from \eqr{g1form}, 
\e{g1sum}{\g^{\1}_{0,\ell>0} = -2\sum_{m=0}^\i R_m(2\D) I_{\ell}(\D,\D'+2m)\,.}
(Recall that the direct-channel amplitude only contributes to $\ell=0$, since it evaluates to a constant on the pole at $2\D$.)

We split the analysis into two parts. First, we evaluate $I_{\ell}(\D,\d)$, i.e. we determine the contribution to the anomalous dimension from a single pole. Next, we perform \eqr{g1sum}, summing over contributions from all poles.

To evaluate $I_{\ell}(\D,\d)$, we close the contour to the left, picking up an infinite series of poles at $s=0,-2,-4,\ldots$. The resulting infinite sums can be regularized using Hurwitz zeta functions. Upon looking at several examples, one infers the following structure for $\D\in\mathbb{Z}$:
\e{}{I_{\ell}(\D,\d) = \sum_{n=n_{min}}^2P_n(\d)\zeta\left(n,{\d\over 2}\right)\,,}
where
\e{}{n_{min} = -(\ell+2\D-4)\,.}
The $P_n(\d)$ are degree-$(n-n_{min})$ polynomials in $\d$, and $P_1(\d)=0$. All but the $\zeta\left(2,{\d\over 2}\right)$ terms reduce to Bernoulli polynomials in $\d$,
\e{}{\zeta\left(-n,{\d\over 2}\right) =-{B_{n+1}({\d\over 2})\over n+1}\,. }
Since $B_{n+1}$ is degree-$(n+1)$, we can rewrite the form of $I_{\ell}$ as
\e{1850}{I_{\ell}(\D,\d) = {\cal P}_{\ell+2\D-3}(\d)+{\cal R}_{\ell+2\D-2}(\d)\zeta\left(2,{\d\over 2}\right)\,,}
where ${\cal P}_m$ and ${\cal R}_m$ are polynomials of degree $m$.\footnote{There is some potential ambiguity in these polynomials; this can be fixed in a given case by comparing to numerical integration.} Note that $\zeta(2,x) = \psi'(x) = d^2_x \log \Gamma(x)$. 

Now we want to sum over all poles at $\d=\D'+2m$. Plugging \eqr{1850} into \eqr{g1sum},
\e{}{\g^{\1}_{0,\ell>0} = -2\sum_{m=0}^\i R_m(2\D) \left({\cal P}_{\ell+2\D-3}(m) + {\cal R}_{\ell+2\D-2}(m)\psi'({\D'+m}) \right)\,,}
where we regard $\D'$ as fixed. The second term is somewhat tricky. To proceed we employ two regularization methods.

The first is an exponential regularization. This is useful when evaluating the sum over ${\cal P}_{\ell+2\D-3}(m)$, e.g. as
\e{}{\sum_{m=0}^\i R_m(2\D){\cal P}_{\ell+2\D-3}(m)e^{-\eps m}\,. }
Performing the sum and expanding near $\eps=0$, the prescription is to keep the finite term, dropping terms that are power law divergent.  

The second is an integral regularization. This is useful when evaluating the sum over ${\cal R}_{\ell+2\D-2}(m)\psi'({\D'+m})$. Specifically, we turn to the integral representation of $\psi'(\D'+m)$,
\e{}{\psi'(\D'+m) = \int_0^\i dt \,{t\,e^{-t(\D'+m)}\over 1-e^{-t}}\,.}
Swapping the order of the sum over $m$ and the integral, performing the sum over $m$, and then performing the integration analytically, the prescription is to keep the finite term, dropping terms that are power law divergent near $t=0$. 

We have checked that these two methods agree in several examples in which both can be carried to the end, e.g. a tree-level scalar exchange with $\D=2,\D'=3$.

\ssec{Application: $\g^{\2}_{0,\ell}$ in $\phi^4$}
We now apply the above to compute $\g^{\2}_{0,\ell}$ for $\ell=2,4$. To make contact with the previous section, we take $d=4, \D=2$, and use the residues \eqr{d4res}. From \eqr{g1sum}, 
\e{g0ls}{\g^{\2}_{0,\ell>0}= -2\sum_{m=0}^\i R_m I_\ell(2,4+2m)\,.}

First we compute $I_{\ell}(2,4+2m)$. Let's first focus on $\ell=2$. Closing the contour in \eqref{mellindelta} to the left, we pick up the poles at $s=0,-2,-4,\ldots$, which yields the following infinite sum:
\e{}{I_2(2,4+2m) = \sum_{k=0}^\i \frac{(k+1) \left(15 k^3+5 k^2 (4 m+13)+k (40 m+86)+22 m+38\right)}{6 (k+m+2)^2}\,.}
We can regularize this using Hurwitz zeta functions of the form $\z(y,m+2)$ for $y=-2,-1,0,2$. Comparing this result to numerical integration, we find that an extra additive polynomial is required. The end result is
\e{I2}{I_2(2,4+2m) \rar {\cal P}(m) +{\cal R}(m)\psi'(m+2)\,,}
where
\es{}{{\cal P}(m) &\equiv {1\over 36}(30m^3+75m^2+71m+23)\,,\\
{\cal R}(m) &\equiv - {1\over 6}(m+1)^2(5m^2+10m+6)\,.}

We now need to perform the sum \eqref{g0ls}. We first do the sum over ${\cal P}(m)$ using an exponential regulator. The sum yields a linear combination of generalized hypergeometric functions; upon expanding in small $\eps$ and keeping the finite term, we get
\es{d4s1}{\sum_{m=0}^\i {\cal P}(m)  R_m  e^{-\eps m} \rar \frac{76}{105}\a^2\,.}
Next, we use the integral regularization on the sum over the ${\cal R}(m)$ term. After performing the sum inside the integral, we have
\es{}{\sum_{m=0}^\i {\cal R}(m)  R_m \psi'(m+2) &= \int_{0}^\i dt {t\over 1-e^{-t}}\left(\frac{3\a^2}{256 \left(e^t-1\right)^{11/2}}\right)\\
&\times \Big(\sqrt{e^t-1} \left(6880 e^t+5848 e^{2 t}+446 e^{3 t}+9 e^{4 t}+992\right)\\&+\arcsin \big(e^{-\frac{t}{2}}\big)\left(4608 e^t+7536 e^{2 t}+1584 e^{3 t}+72 e^{4 t}-9 e^{5 t}+384\right) \Big)\,.}
Performing the integral, and keeping the finite terms,
\e{d4s2}{\sum_{m=0}^\i {\cal R}(m)  R_m \psi'(m+2) \rar -\frac{583+174 \pi ^2}{3465}\a^2\,.}
Adding this to \eqr{d4s1} and multiplying by (-2) to obtain \eqr{g0ls}, the final result is
\e{}{\g^{\2}_{0,2} =\frac{2(174\pi^2-1925)}{3465} \alpha^2\,.}
This agrees with equation \eqref{d4l2}. 

An analogous procedure can be carried out for $\ell=4$. The analog of \eqr{I2} is 
\e{I4}{I_4(2,4+2m) \rar {\cal P}(m) + {\cal R}(m) \zeta(2,m+2)\,,}
with
\es{}{{\cal P}(m) &= \frac{630 m^5+2835 m^4+6195 m^3+7350 m^2+4579 m+1159}{1800}\,,\\
{\cal R}(m) &= -\frac{1}{60} (m+1)^2 \left(21 m^4+84 m^3+161 m^2+154 m+60\right)\,.}
Carrying out the sum over $m$ using the above techniques, we get
\es{}{\sum_{m=0}^\i {\cal P}(m)  R_m \rar \frac{104267}{214500}\a^2~, \quad \sum_{m=0}^\i {\cal R}(m)  R_m \psi'(m+2)  \rar -\frac{2 \left(83636+18825 \pi ^2\right)}{1126125}\a^2\,.}
Adding the two numbers and multiplying by (-2), this agrees with \eqr{d4l4}. We have repeated all of the above for $d=2$, finding agreement there as well.

\vs
To summarize, the results of this subsection give further confirmation that our solution to the crossing problem is equivalent to a direct computation of $\phi^4$ one-loop Witten diagrams in AdS. We reiterate that this agreement acts as a check on a match between two independent techniques used to derive anomalous dimensions: on the one hand, the large spin resummation technique used in the crossing problem, without reference to any amplitude; and on the other, the techniques of this section used to extract low-spin data from $\mloop$. 

\ssec{A remark on $\phi^3$ theory}
There is a small subtlety when computing $\g^{\1}_{n,\ell}$ for $\D'=\D$. A common example is for the leading pole of an exchange diagram in $\phi^3$ theory. The fully symmetrized contribution of a pole at twist $\D$ in all three channels is 
\e{}{{1\over s-\D}+{1\over t-\D}+{1\over \hat u-\D}~.}
If we evaluate the $s$ and $\hat u$ poles on the $\G^2$ double-trace pole at $t=2\D$, they cancel. Thus, naively, so do the $s$- and $\hat u$-channel contributions to $\g^{\1}_{n,\ell}$. However, they are supposed to add, just as they do for $\d\neq \D$. To get around this, one can simply deform the internal dimension by a small amount, $\d=\D+\eps$, perform the computation in which the two channels add, and then take $\eps\rar 0$. A spacetime computation confirms this result. For example, see Appendix \ref{appd} for the computation of $\g^{\1}_{n,\ell}$ in $\phi^3$ theory for $\D=2$.

\sec{Discussion}\label{s7}

In this paper we initiated an analysis of large $N$ CFT four-point correlators at next-to-leading order in $1/N$, which map by the AdS/CFT correspondence to one-loop diagrams in AdS space. We presented general methods to analyze correlation functions at this order, and implemented them explicitly for two examples: a $\phi^4$ theory in the bulk, and a $\phi^3+\phi^4$ theory in the bulk.

There are various levels of extension of what we have done here, most of which are needed in order to study the $1/N$ expansion in more generic, full-fledged holographic CFTs. We first discuss some of these, and then move on to broader future directions. 
\ssec{Generalizations}

\begin{itemize}

\item An immediate priority, and a necessary step toward solving bona fide CFTs, is to solve the crossing equations when the OPEs contain single-trace operators. In the single-scalar theory, this would yield a computation of the scalar box diagram in AdS $\phi^3$ theory. 

\item We could also allow exchanges of operators with spin. These should present some technical complications, but we do not expect them to lead to any qualitative changes. A particularly important version of this is to incorporate the stress tensor, which allows us to access graviton loops.\footnote{Note that for operators of $\Delta\in\mathbb{Z}$, there is potential mixing between $[\cO \cO]$ and $[T T]$, at least in some channel where global symmetry-singlet $[\cO \cO]$ operators contribute.}

\item One would also like to extend our methods to include multiple species of operators, as in \cite{Heemskerk:2010ty}. Indeed, a generic CFT, as opposed to a bottom-up generalized free field theory, always has an infinite number of single-trace operators. If there are additional fields $\chi$ in the bulk with four-point couplings ${\tilde \mu}_4 \phi^2 \chi^2$, then the corresponding bubble diagrams are also easy to compute, given the tree-level $\langle \cO \cO \cO_{\chi} \cO_{\chi} \rangle$; see \cite{Fitzpatrick:2011hu}. When there are also three-point $\phi \chi \chi$ vertices, the situation is more complicated. The case when some $[\cO_{\chi} \cO_{\chi}]$ operator is degenerate with a $[\cO \cO]$ operator with the same quantum numbers is discussed in Appendix \ref{degen}, and requires generalizing the bootstrap analysis to different external operators. It should be straightforward to understand the form of $\mloop$ when external dimensions are unequal. The basic structure will be identical to \eqr{mloopgen}: $\mloop$ will have poles at $\t=2\D+2n$ in every channel, with residues fixed by first-order data.

\item An extension to higher loops would also be profitable. On the CFT side, one will generally have to contend with triple- and higher-trace operators. By the arguments of Section \ref{s3}, one sees that at $O(1/N^{2(L+1)})$ -- dual to $L$-loop order in AdS -- $M_{L-{\rm loop}}$ has poles of degree $\leq L$. In general theories, going to $O(1/N^6)$ requires additional information from five-point functions. However, in theories that have a $\cO \leftrightarrow -\cO$ symmetry, these five-point functions vanish (and correspondingly no triple-trace operators appear in the $\cO\times \cO$ OPE). So in these theories it may be possible to compute the four-point functions and anomalous dimensions also at two-loop order, with no new conceptual wrinkles. 

What we really seek, however, are the AdS loop-level Feynman rules for Mellin amplitudes, thus giving an algorithm for any $L$-loop calculation.

\item In general the $1/N$ expansion is only asymptotic, and there are non-perturbative effects scaling as (say) $e^{-N}$ that must be understood before even attempting to continue the large $N$ results to finite values of $N$. Can we use our methods also for such non-perturbative contributions?

\item 
As we discussed, the crossing analysis simplifies considerably when the dimension $\Delta$ is an integer. It would be interesting to analyze the case of non-integer $\Delta$ and to obtain explicit results for this case as well.

\item For theories with three-point vertices, we found (see \eqr{huv}) that the four-point function at $O(1/N^4)$ has a contribution proportional to $u^{\Delta} \log^2(u) \log^2(v)$, with a crossing-symmetric coefficient function $h(u,v)$. What are the form and content of this function?

\item
Our analysis in this paper did not assume any additional symmetries. It should be simple to take into account additional global symmetries. Incorporating supersymmetry should also be straightforward, at least in principle, with superconformal blocks replacing the conformal blocks. 

An especially interesting example, as always, is the $d=4$, ${\cal N}=4$ SYM theory. In the $\lambda_{YM} = g_{YM}^2 N \to \infty$ limit, the bulk theory only contains the fields dual to protected single-trace operators. The solution to crossing in this limit at order $1/N^2$ was performed in \cite{Alday:2014tsa}. Its generalization to order $1/N^4$ involves all the issues mentioned earlier: in particular, there is an infinite number of single-trace operators, and they all have integer dimensions so that there can be complicated mixings between the various double-trace operators. Luckily, the four-point functions of all these protected single-trace operators were recently computed in \cite{Rastelli:2016nze}, and this information should be sufficient to work out the mixing matrix, and thus to compute the correlation functions of protected operators in this theory at order $1/N^4$. It would be interesting to perform this analysis. 
  
In fact, the analysis should be simpler than it may appear. We now make a potentially powerful observation. Consider the Mellin amplitude for the four-point function $\la \twp \twp \twp \twp\ra$. At large $\l_{YM}$, all single-trace operators in the $\twp\times\twp$ OPE have even twist: these are the operators $\O_k$ with even $k\geq 2$, which are the 1/2-BPS operators in the $[0,k,0]$ representation of $SU(4)$. ($\twp \equiv \O_2$.) Due to non-renormalization of the $\O_k$ dimensions and $SU(4)$ selection rules, the only operators appearing in $\mloop$ will be the double-trace operators $[\O_k\O_k]_{n,\ell}$, with $k\geq 2$; so all poles in $\mloop$ sit at even twist $\t= 4+2n$. Now, for every value of $n$, there are double-trace operators $[\twp\twp]_{n,\ell}$. If we compute $\mloop^{\twp\twp}$ for this correlator -- that is, the piece of $\mloop$ fixed by requiring a match to the contributions of the $[\twp\twp]_{n,\ell}$ operators, as done in Section \ref{s3} -- the residues at {\it all} twists $\t= 4+2n$ are fixed by $[\twp\twp]_{n,\ell}$ tree-level data. Therefore, up to regular terms, this fixes the {\it full} one-loop amplitude! That is, up to regular terms, 
\e{meq}{\mloop = \mloop^{[\twp\twp]}~.}
While the operator mixings mentioned above still plague the calculation, \eqr{meq} says that the full one-loop amplitude -- which involves an infinite set of diagrams involving virtual $\O_k$ loops -- is determined just by the anomalous dimensions $\g^{\1}_{n,\ell}$ for the $[\twp\twp]_{n,\ell}$ operators. This is a great simplification, apparently due to the effect of maximal supersymmetry on the spectrum.

In general, for a four-point function of some operator $\O$, this coincidence of poles occurs whenever $\D_{\O}\in\mathbb{Z}$ and the spectrum of twists in the $\O\times\O$ OPE is even.\footnote{This phenomenon has a tree-level version: in a tree-level exchange of twist $\d$ between external operators of dimension $\D$, the amplitude has only a finite number of poles when $\d-2\D\in2\mathbb{Z}$. This happens because the single-trace and double-trace poles collide, and would thus produce a triple pole, violating the $1/N$ expansion, unless these single-trace poles drop out of the amplitude. This was also recently noted in \cite{Rastelli:2016nze}.} Besides ${\cal N}=4$ SYM, this also occurs when $\O$ is the bottom component of the stress tensor multiplet of  the $d=6$, ${\cal N}=(2,0)$ theory of M5-branes. We can also analyze many other interesting supersymmetric conformal field theories, such as the $d=3$, ${\cal N}=8$ theory of M2-branes (which does not have the same simplification described above). In both of these cases, there is again a gap to the non-protected operators, but here it scales as a power of $N$ that does not involve an extra independent parameter. Thus one cannot separate the loop expansion and the derivative expansion in the bulk.\footnote{At some specific low orders in $1/N$, it is possible to separate the different contributions to the correlation functions: in particular, the leading $1/N$ correction is due not to a loop, but to a higher-derivative correction to the action that descends from anomalies in $d=11$ supergravity (e.g. \cite{Tseytlin:2000sf}).} In any case, the loop diagrams in the dual AdS bulk can still be computed by the methods described in this paper.

\end{itemize}

\ssec{Future directions}

\begin{itemize}

\item When we have a standard field theory in AdS space (as opposed to a gravitational one), it has not just correlation functions with sources at the boundary as we discussed in this paper, but also correlation functions of operators at arbitrary bulk points. Are these determined in terms of the correlation functions with boundary sources? Can we say anything about them by our methods?

\item In our discussion of the ${\cal N}=4$ SYM theory we integrated out the stringy states, but we can repeat the same story when including non-protected string states. For finite $\lambda_{YM}$ there are additional operators contributing with dimensions at least of order $\lambda_{YM}^{1/4}$. These operators can be integrated out in an expansion in $1/\lambda_{YM}^{1/4}$, whose form at order $1/N^2$ was discussed in \cite{Alday:2014tsa}. Using this information it should be possible to work out also the order $1/N^4$ correlators in a systematic expansion in $1/\lambda_{YM}^{1/4}$. Can we use the large $N$ expansion of the crossing equation to learn anything about the non-protected states?

\item For bulk theories which are string theory backgrounds, the $1/N$ expansion (the loop expansion in the bulk) coincides with the genus expansion of the worldsheet theory. The correlators we discuss arise as integrated correlation functions in this worldsheet theory. What does our analysis teach us about these worldsheet theories? Can we relate the crossing equations in the CFT and in the worldsheet theory?

\item We close with some words on the relation of the large $N$ bootstrap to flat space physics. An alternative way to approach the AdS amplitudes problem might have been to start from known facts about S-matrices, and find analogs or extensions to AdS. We took a different tack, but it would be very interesting to turn to these questions using our results. In \cite{Fitzpatrick:2011hu}, the emergence of the optical theorem in the flat space limit was studied, but one would also like to know whether there is a direct analog at finite AdS curvature. 

Similarly, in our one-loop crossing computations we found specific harmonic polylogarithms appearing. As we noted, this suggests an intriguing underlying structure akin to flat space amplitudes. On the other hand, the one-loop Mellin amplitudes themselves were given by the more familiar generalized hypergeometric functions and, in the case of \eqr{mlooptrid2}, a digamma function. What class of functions forms a basis for the multi-loop solution of the crossing equations, and for the AdS Mellin amplitudes themselves? Which diagrams form a basis for all others at a given loop order? The answers would presumably be closely related to the possible existence of AdS analogs of generalized unitarity, on-shell methods and the like. It would be fascinating to try to understand the big picture here.

Finally, we note that Mellin amplitudes admit flat space limits \cite{Penedones:2010ue}. If one can develop the solution to crossing to successively higher orders in $1/N$, taking that limit would shed light on flat space higher-loop amplitudes. A specific, and difficult, longer-term challenge in the supergravity community is to determine the critical dimension above which the four-point, five-loop amplitude in maximal supergravity diverges. This has resisted years of direct attack using advanced methods \cite{Bjornsson:2010wm, Bern:2012uf, Kallosh:2014hga}. It would be fascinating if, eventually, the five-loop crossing equations, applied to the holographic dual of gauged maximal supergravity, could be employed in this endeavor.

\end{itemize}

\section*{Acknowledgments}

We wish to thank Nima Afkhami-Jeddi, Michael B. Green, Vasco Goncalves, Tom Hartman, Zohar Komargodski, David Simmons-Duffin, Ellis Ye Yuan and Sasha Zhiboedov for helpful discussions. The work of OA was supported in part  by the I-CORE program of the Planning and Budgeting Committee and the Israel Science Foundation (grant number 1937/12), by an Israel Science Foundation center for excellence grant, by the Minerva foundation with funding from the Federal German Ministry for Education and Research, by a Henri Gutwirth award from the Henri Gutwirth Fund for the Promotion of Research, and by the ISF within the ISF-UGC joint research program framework (grant no.\ 1200/14). OA is the Samuel Sebba Professorial Chair of Pure and Applied Physics. The work of LFA was supported by ERC STG grant 306260. LFA is a Wolfson Royal Society Research Merit Award holder. AB acknowledges the University of Oxford for hospitality where part of this work has been done. AB is partially supported by Templeton Award 52476 of A. Strominger and by Simons Investigator Award from the Simons Foundation of X. Yin. EP is supported by the Department of Energy under Grant No. DE-FG02-91ER40671. 

\appendix

\sec{Operator content of the one-loop crossing equations}
\label{crossing_ops}

In this appendix we discuss the operators that can appear in the OPE of two identical single-trace primary operators $\cO$ and $\cO$ of dimension $\Delta$, and at which order in a large $N$ expansion they contribute to the crossing equation. The upshot is that at order $1/N^4$, we do not have to consider any operators with more than two traces appearing in the OPE. 

The notation is that $[\cO_1 \cO_2 ... \cO_m]$ is an $m$-trace primary operator corresponding to an $m$-particle state in the bulk (and appearing at $N=\infty$ in the OPE of $\cO_1(x_1) \cO_2(x_2) \cdots \cO_m(x_m)$; for the precise definition at $m=2$, see appendices of \cite{Penedones:2010ue,Fitzpatrick:2011dm}). All operators will be normalized such that their two-point function is one. We will choose a basis in which there is no mixing between operators with a different number of traces (we will discuss mixings of different double-trace operators below). This means, for instance, that $[\cO_i \cO_j]$ is not exactly the operator appearing in the OPE of $\cO_i$ and $\cO_j$, but may differ from it at order $1/N$; these differences will not be important in the order we work in.

On general grounds, connected $n$-point functions of single-trace operators scale as $1/N^{n-2}$ in the large $N$ limit. Naively this implies that the OPE coefficient of a $k$-trace operator, proportional to $\langle \cO \cO [\cO_1 \cdots \cO_k] \rangle$ scales as $1/N^{k}$. In general this expectation can fail only if there is an extra disconnected contribution to this correlation function. However in our case, since we chose the single-trace operators to be orthogonal to operators with more traces, such a disconnected correlation function can only appear for the operators $[\cO \cO]$, which have OPE coefficients of order one. Thus the OPE coefficient of operators with three or more traces is suppressed at least by $1/N^3$, so they will not contribute to the crossing equation at order $1/N^4$.

The only operators contributing at order $1/N^4$ are then:

\begin{itemize}

\item Single-trace operators $\cO_1$, with some even spin $\ell$ (a special case is the energy-momentum tensor): The OPE coefficient $c_{\cO \cO \cO_1}$ is generically of order $1/N$, so they contribute to crossing already at order $1/N^2$. At order $1/N^4$ we will see corrections to these contributions due to $1/N^2$ corrections to the dimensions of $\cO$ and $\cO_1$, and to $c_{\cO \cO \cO_1}$. These cannot be determined by crossing since they are the basic inputs -- in the bulk these are masses and three-point vertices that need to be determined by some renormalization condition at all orders in $1/N$. Thus from the point of view of the crossing equation we need to take these as given. If we use renormalization conditions that are independent of $N$, and in particular for protected operators in superconformal field theories, single-trace operators will appear in the crossing equation {\it only} at order $1/N^2$; otherwise their contributions at higher orders are simply related to the leading order contribution and to the corrections to the dimensions and single-trace OPE coefficients.

\item Double-trace operators $[\cO \cO]_{n,\ell}$ : These appear already at order $1$ with squared OPE coefficients $a_{n,\ell}^{(0)}$, and with dimensions $2\Delta+2n+\ell$. As we discuss extensively, at higher orders in $1/N$ they give contributions related to the corrections to the OPE coefficients and dimensions of these double-trace operators.

\item Other double-trace operators $[\cO_1 \cO_2]_{n,l}$ : The OPE coefficient $c_{\cO \cO [\cO_1 \cO_2]_{n,l}}$ is of order $1/N^2$. Thus, generally these operators appear in the crossing equation at order $1/N^4$, with a contribution depending on the leading order dimension $\Delta_1+\Delta_2$, and on the leading order $c_{\cO \cO [\cO_1 \cO_2]_{n,l}}$. The latter depends on four-point couplings in the bulk which are arbitrary, so from the point of view of the four-point function $\langle \cO \cO \cO \cO \rangle$ they will give us parameters that we cannot determine. However, because these contributions depend only on the leading order dimensions, they generically do not come with any logs in the direct channel, so they will not affect the universal terms that we discuss in this paper; they give rise to independent poles in Mellin space. This is not true when these operators mix with the $[\cO \cO]$ operators, as we discuss below.

\end{itemize}

At order $1/N^6$ the analysis will change, and in particular triple-trace operators will also start appearing, depending on (undetermined from crossing) five-point vertices in the bulk.

\subsec{Degeneracies} 
\label{degen}

One important issue that was ignored  in the analysis above is mixing between different double-trace operators when they are degenerate; this often happens in interesting examples, and a mixing of $[\cO \cO]$ with other double-trace operators significantly modifies the analysis.

As a typical example, consider a $\phi^2 \phi_1^2$ field theory on AdS, where $\phi$ and $\phi_1$ are scalars with the same mass, and where $\cO$ is dual to $\phi$ and $\cO_1$ to $\phi_1$. In this theory, the OPE of $\cO$ and $\cO$ contains $[\cO \cO]$ starting at order $1$ from the disconnected diagram in AdS, and $[\cO_1 \cO_1]$ starting at order $1/N^2$ from an X-shaped diagram, and no single-trace operators. There are no tree-level diagrams contributing to $\langle \cO \cO \cO \cO \rangle$ and to $\langle \cO_1 \cO_1 \cO_1 \cO_1 \rangle$, so naively the analysis at order $1/N^2$ implies that $[\cO \cO]$ and $[\cO_1 \cO_1]$ have no anomalous dimensions at this order. We then expect  to have no logarithmic terms in the direct-channel four-point function at order $1/N^2$, and no double-logs at order $1/N^4$ (i.e. no poles in $M_{\rm 1-loop}$). But on the other hand, the one-loop diagram contributing to $\langle \cO \cO \cO \cO \rangle$ is clearly the same as in the $\phi^4$ theory, which does have such double-logs/poles since the latter theory does have a non-trivial tree-level diagram.

The resolution is that the two double-trace operators mix: there is a bulk tree-level diagram giving a non-zero $\langle [\cO \cO] [\cO_1 \cO_1] \rangle \sim 1/N^2$. The correct basis of operators with diagonal two-point functions is $\cO_A \equiv [\cO \cO] + [\cO_1 \cO_1]$ and $\cO_B \equiv [\cO \cO] - [\cO_1 \cO_1]$, and it turns out that the first operator has an anomalous dimension $\gamma_A^{(1)}=C/N^2$ and the second operator $\gamma_B^{(1)}=-C/N^2$, for some constant $C$. This reproduces the two-point functions at order $1/N^2$.
It also explains why we get double-logs at order $1/N^4$ (poles in $M_{\rm 1-loop}$) in $\langle \cO \cO \cO \cO \rangle$, since these are proportional to $(\gamma_A^{(1)})^2 + (\gamma_B^{(1)})^2$ (both $\cO_A$ and $\cO_B$ appear in the $\cO \times\cO$ OPE), which is non-zero.

The lesson is that in general, we have to be careful of double-trace mixings; all operators that mix with $[\cO \cO]_{n,\ell}$ appear in the crossing equation already at order $1/N^2$, and will lead to double-logs at order $1/N^4$. The coefficients of these double-logs cannot be computed without knowing the precise mixing matrix: one has to know all correlators $\langle [\cO \cO] [\O_1 \O_2] \rangle$ at order $1/N^2$, which can be extracted from tree-level $\langle \cO \cO \cO_1 \cO_2 \rangle$ four-point functions, before one can use the crossing equation at order $1/N^4$. Note that mixings of this type occur in the ${\cal N}=4$ SYM theory, complicating its analysis.

\section{Explicit expansions}\label{appb}
In this appendix we display explicit results for the large spin expansion for $\gamma_{0,\ell}^{(2)}$ in several examples. Recall that the total result is the sum over contributions from each conformal block in the dual channel. In a case in which the solution at order $1/N^2$ has support only for spin zero we obtain
\begin{equation}
\label{finalgamma2}
\gamma_{0,\ell}^{(2)}= \frac{1}{8} \sum_{n} a_{n,0}^{(0)} \left(\gamma_{n,0}^{(1)} \right)^2 \left. \gamma_{0,\ell}^{(2)} \right|_{(n,0)},~~~~~~~~\left. \gamma_{0,\ell}^{(2)} \right|_{(n,s)}=-\frac{c^{(0)}_{n,s}}{J^{2\Delta+2n}} \left. \hat \gamma_{0,\ell}^{(2)} \right|_{(n,s)}~.
\end{equation}
$\left.\hat  \gamma_{0,\ell}^{(2)} \right|_{(n,0)}$ has the structure explained in Section \ref{s4}. 

For $\Delta=2$ in $d=4$ we obtain for the first few cases
\begin{eqnarray}
\left.\hat  \gamma_{0,\ell}^{(2)} \right|_{(0,0)}&=&\frac{2 (\ell+2) (\ell (\ell+4)+5)}{\ell+1}+2J^4  \psi ^{(2)}(\ell+1)\,,\\
\left.\hat  \gamma_{0,\ell}^{(2)} \right|_{(1,0)} &=& \frac{3}{2} (\ell+2)^2 (\ell (\ell (6 \ell (\ell+7)+115)+148)+77)+ 3 J^6 \left(3 \ell^2+9 \ell+8\right) \psi ^{(2)}(\ell+1) \nonumber\,,\\
\left.\hat  \gamma_{0,\ell}^{(2)} \right|_{(2,0)} &=&\frac{5}{36} (\ell+1) (\ell+2)^3 (\ell (\ell (3 \ell (\ell (10 \ell (\ell+10)+443)+1111)+4975)+4187)+1558) \nonumber \\
& & +\frac{5}{6} J^8 \left(5 \ell^4+30 \ell^3+79 \ell^2+102 \ell+54\right) \psi ^{(2)}(\ell+1)\,, \nonumber
\end{eqnarray}
where we have introduced $J^2=(\ell+1)(\ell+2)$. As explained in the body of the paper, with these ingredients it is possible to obtain $\gamma_{0,\ell}^{(2)}$ also for finite values of the spin $\ell$. For spin zero, we get
\begin{align}
\gamma_{0,0}^{(2)}=\sum_{n}\left(\frac{36(1+n)^4(5+2n(3+n))}{(3+4n(2+n))}+\frac{72(1+n)^8}{3+4n(2+n)}\psi ^{(2)}(n+1)\right)\alpha^2\,.
\end{align}
As already discussed, this sum is divergent.

For $\Delta=2$ in $d=2$, equivalently one can find the form of $\hat \gamma_{0,\ell}^{(2)} |_{(n,0)}$ and compute $\gamma_{0,\ell}^{(2)}$. For   spin zero in this case we obtain
\begin{align}
\gamma_{0,0}^{(2)}=\sum_{n}\left(\frac{9(19+n(25+2n(6+n)))}{2(1+n)}+\left(9(1+n)^2(2+n)^2\right)\psi ^{(2)}(n+1)\right)\alpha^2=\frac{15-4 \pi^2}{10}\alpha^2\,.
\end{align}
For each model we can obtain the expansion of $\gamma_{0,\ell}^{(2)}$ around large $\ell$. The expansion is better organised in powers of $J^2$. For instance, for the interaction $\phi^4$ with $\Delta=2$ in $d=4$ we obtain
\begin{equation}
\gamma_{0,\ell}^{(2)} = -\frac{12}{J^4}\left(1+\frac{18}{5} \frac{1}{J^2}+\frac{96}{7} \frac{1}{J^4}+\frac{360}{7} \frac{1}{J^6} +\frac{74304}{385} \frac{1}{J^8}+\frac{724320}{1001}\frac{1}{J^{10}}+\cdots \right)\alpha^2 \,,
\end{equation}
for the interaction $\phi^4$ with $\Delta=2$ in $d=2$ we obtain
\begin{equation}
\gamma_{0,\ell}^{(2)} = -\frac{6}{J^4}\left(1+\frac{4}{5} \frac{1}{J^2}+\frac{4}{7} \frac{1}{J^4}+\frac{16}{35} \frac{1}{J^6} +\frac{16}{55} \frac{1}{J^8}+\frac{1856}{5005}\frac{1}{J^{10}}+\cdots \right)\alpha^2 \,,
\end{equation}
while for the mixed interaction ${\mu_3\over 3!}\phi^3+{\mu_4\over 4!}\phi^4$ with $\Delta=2$ in $d=4$ we obtain
\begin{equation}
\gamma_{0,\ell}^{(2)} = \frac{80}{J^4}\left(1+\frac{18}{25} \frac{1}{J^2}+\frac{96}{175} \frac{1}{J^4}+\frac{72}{175}\frac{1}{J^6} +\frac{576}{1925} \frac{1}{J^8}+\frac{6624}{25025}\frac{1}{J^{10}}+\cdots \right)\mu_3^2 \mu_4 \,.
\end{equation}

The expansions above are asymptotic. In the body of the paper we have shown how to resum the expansions and compute them for finite values of the spin. It is interesting to compare the asymptotic series above with the correct results for different values of the spin. For instance, for $\ell=2$ we have $J^2=12$. Including the first six terms shown above for $\phi^4$ and $\phi^3+\phi^4$ in $d=4$ we would obtain
\begin{eqnarray}
\gamma_{0,2}^{(2)} &\approx& -0.11977 \alpha^2,~~~~\text{for $\phi^4$}\,,\\
\gamma_{0,2}^{(2)} &\approx& 0.591144 \mu_3^2 \mu_4,~~~~\text{for $\phi^3 + \phi^4$}\,,
\end{eqnarray}
to be compared with the exact values
\begin{eqnarray}
\gamma_{0,2}^{(2)} &=&\frac{2}{3465}(174\pi^2-1925) \alpha^2 \approx -0.11988 \alpha^2,~~~~\text{for $\phi^4$}\,,\\
\gamma_{0,2}^{(2)} &=&\frac{16}{945}(39\pi^2-350)  \mu_3^2 \mu_4 \approx 0.591146 \mu_3^2 \mu_4,~~~~\text{for $\phi^3 + \phi^4$}\,.
\end{eqnarray}
We see that the values we obtain from the asymptotic series are remarkably close to the correct values, even for spin two! In the case of convergent answers, even the approximation for spin zero is very good.

\section{General expectations for UV divergences and the large $n$ limit of $\gamma^{(2)}_{n,\ell}$}
\label{UVdiv}

When we compute bulk loop diagrams we expect to get UV divergences. Since these arise at short distances, they should take a similar form in AdS as in flat space, and at any loop order we should be able to cancel them by local counter-terms in AdS. In general the bulk theories we discuss are effective theories which are non-renormalizable, so they require a cutoff, and at higher orders in perturbation theory we will need to add more and more counter-terms, but in this paper we just discuss the one-loop order. As argued in the Introduction, in our bootstrap computation related to a divergent bulk diagram we expect to find a divergence in $\gamma^{(2)}_{n,\ell}$, and we expect that when we regularize it (for instance by putting some cutoff on the sums), the divergence is precisely proportional to $\gamma^{(1)}_{n,\ell}$ coming from some local bulk terms, so that it can be removed by putting in appropriate cutoff-dependent bulk terms. 

Recall that on general grounds we expect any local bulk term that is allowed by the symmetries to appear with an arbitrary coefficient, both from the bulk point of view, and from the bootstrap point of view, since any such term gives a solution to the crossing equations. Thus at any loop order any solution that we find for the four-point function, both from the field theory and bootstrap points of view, is just up to bulk terms. This means that we should take both the dimensions and the three-point functions of single-trace operators, at all orders in $1/N$, to be inputs to the computation, that we cannot determine just from the crossing equations in a $1/N$ expansion. In addition we have a freedom to choose any local four-point terms, namely to shift the solution by any of the ``homogeneous'' solutions to the crossing equations that correspond to finite-order polynomials in Mellin space (we called them $f_{\rm reg}$ in Section \ref{s331}). We expect to need this freedom in order to cancel divergences. We cannot fix it just from crossing. 

Consider first the $\lambda \phi^4$ theory in $AdS_5$. The coupling constant $\lambda$ here has dimensions of length, and one can define a dimensionless coupling $\lambda/R_{AdS}$, that in our $1/N$ expansion is proportional to $1/N^2$.

In flat space the four-particle tree-level scattering amplitude goes like $\lambda$; when we translate it into some dimensionless quantity this will go at high energies as $\lambda E$ where $E$ is a typical energy. In AdS the role of the energy is played by $n$, so we expect to find for the tree-level four-point amplitude a result going as 
\begin{equation}
\gamma^{(1)}_{n,\ell} \propto \lambda n \simeq {n\over N^2}
\end{equation}
at large $n$, which is indeed what we find \eqref{gammaphi4}. (In this case the answer happens to vanish for $l>0$.) Note that large $n$ here means $n \gg 1$ and $n \gg \Delta$, so that the energy is larger than the mass and the scale of the AdS radius.

At one-loop in flat space we have a linear divergence, and the amplitude with a finite cutoff $\Lambda$ goes at high energies as $\lambda^2 (\Lambda + E + \cdots)$. Note that we do not get a logarithmic divergence; indeed such a divergence would multiply $E$ but there is no local counter-term that could cancel this (higher-derivative couplings in the bulk give higher powers of $E$). Noting that the divergence is just a constant, it can be canceled by shifting $\lambda$ by a term proportional to $\lambda^2 \Lambda$. Translating to AdS as above, we expect to find for the one-loop, four-point function at large $n$
\begin{equation}
\gamma^{(2)}_{n,\ell} \propto \lambda^2 (n^2 + {\tilde \Lambda} n) \simeq {n^2 + {\tilde \Lambda} n\over N^4}~,
\end{equation}
where ${\tilde \Lambda}$ is some cutoff that we use to obtain a finite result. This is a prediction for the large $n$ behavior of $\g^{\2}_{n,\ell}$ in $\phi^4$.
We expect from the locality of the divergence that we could obtain a finite result by shifting $\gamma^{(2)}_{n,\ell}$ by a term proportional to ${\tilde \Lambda} \gamma^{(1)}_{n,\ell}$ of the $\phi^4$ theory. Note in particular that this means that only $\ell=0$ terms should diverge, and this is indeed what we find in Section \ref{s5}. Note also that as far as the crossing equations in the $1/N$ expansion are concerned, there is no obvious way to fix the finite local $\phi^4$ bulk term remaining after this subtraction.

In general we get precise predictions for which divergences we should get in our computation. It should always be possible to cancel divergences in $\gamma^{(k)}_{n,\ell}$ that are related to 4-$\phi$ counter-terms by adding terms proportional to the $\gamma^{(1)}_{n,\ell}$'s that are associated with the counter-terms we need in the bulk. In Mellin space these divergences should always be a polynomial, of a finite degree related to the loop order. Above one-loop, divergences related to counter-terms with more $\phi$'s can also appear.

The analysis of the ${\cal N}=4$ SYM theory, and the related supergravity on $AdS_5$, is analogous. The only difference is that we have to be careful if we regularize our computation, that the regularization preserves supersymmetry, otherwise we will get divergences that are related to bulk counter-terms that are different from the supersymmetric local terms in the bulk. Using a supersymmetric regularization the divergences should all be proportional to the tree-level contributions analyzed in \cite{Alday:2014tsa}. We leave a detailed discussion of this case to the future.

Finally, if we consider the $\phi^3$ theory in $AdS_5$ ($d=4$), the theory is super-renormalizable so there should be no divergences in the four-point functions that cannot be swallowed into the masses and three-point couplings in the bulk. In this case dimensional analysis implies that all $\gamma^{(k)}_{n,\ell}$ should not grow at large $n$, and that we would not encounter any UV divergences in their computation. For $d > 5$ the bulk theory is non-renormalizable, so we expect the large $n$ behavior of the tree-level terms to go as $n^{d-5}$, and one-loop terms to go as the square of this.

\sec{$\phi^3$ OPE data}\label{appd}
In this appendix we derive the tree-level anomalous dimensions of double trace operators due to a fully symmetric exchange of a scalar operator of $\D=2$, i.e. for a $\D=2$ scalar with a ${\mu_3\over 3!}\phi^3$ coupling in AdS. This result was quoted in \eqr{phi3g1n0} for $\ell=0$. 

Such an exchange has been considered in \cite{D'Hoker:1999ni, Penedones:2010ue} and it can be reduced to
\begin{equation}
G(u,v)=\mu(d) u^2 \bar{D}_{1212}(u,v)\,,
\end{equation}
where $\mu(d)$ is a constant which depends on the number of space time dimensions, in particular $\mu(4)=8 \mu_3^2 = 2C_{\O\O\O}^2$ and $\mu(2)=2 \mu_3^2=C_{\O\O\O}^2$. The functions $\bar{D}(u,v)$ are defined as 
\begin{equation}
\bar D_{\Delta_i}(u,v)=\frac{2\prod_i\Gamma(\Delta_i)}{\Gamma(\frac{1}{2}\sum_i\Delta_i-2)}x_{13}^{2\Delta_1}x_{24}^{2\Delta_2}\left(\frac{x_{14}^2}{x_{13}^2x_{34}^2}\right)^{\frac{\Delta_1-\Delta_3}{2}}\left(\frac{x_{13}^2}{x_{14}^2 x_{34}^2}\right)^{\frac{\Delta_2-\Delta_4}{2}}D_{\Delta_i}(x_i)\,,
\end{equation}
where 
\begin{equation}
D_{\Delta_i}(x_i)=\frac{\Gamma(\frac{1}{2}\sum_i\Delta_i-2)}{\prod_i\Gamma(\Delta_i)}\int_{0}^{\infty}\prod_i dt_i t_i^{\Delta_i-1}e^{-\frac{1}{2}\sum_{i,j}t_i t_j x_{ij}^2}\,.
\end{equation}
The symmetry properties of  $\bar{D}(u,v)$ are
\begin{align}
\bar{D}_{\Delta_1 \Delta_2 \Delta_3 \Delta_4 }(u,v)&=v^{-\Delta_2} \bar{D}_{\Delta_1 \Delta_2 \Delta_4 \Delta_3 }(\frac{u}{v},\frac{1}{v})\\
&=\bar{D}_{\Delta_3 \Delta_2 \Delta_1 \Delta_4 }(v,u)\\
&=u^{-\Delta_2}\bar{D}_{\Delta_4 \Delta_2 \Delta_3 \Delta_1 }(\frac{1}{u},\frac{v}{u})\,.
\end{align}
We would like to study the fully symmetrized amplitude which corresponds to 
\begin{align} \label{dbarfour}
\mathcal{G}(u,v)&=G(u,v)+ G\left(\frac{u}{v},\frac{1}{v} \right)+u^2 G\left( \frac{1}{u},\frac{v}{u}\right)\\
&=\mu(d) u^2 \left( \bar{D}_{1212}(u,v)+\bar{D}_{2211}(u,v)+\bar{D}_{1221}(u,v)\right)\,, \label{dbarfour}
\end{align}
where in the last line the symmetry properties of $\bar{D}(u,v)$ have been used. To compute the anomalous dimension it is enough to focus on the terms proportional to $\log u$ in \eqref{dbarfour} and perform the conformal partial wave expansion
\begin{equation} \label{cpwa_anyd}
\mathcal{G}(u,v)|_{\log u}=\frac{1}{2}\sum_{n, \ell} a^{(0)}_{n, \ell} \g^{\1, \phi^3}_{n,\ell} u^{n+2} g_{4+2n+\ell, \ell}(u,v)\,.
\end{equation}
Notice that the small $u$ expansion of $\mathcal{G}(u,v)$ starts at order $u$, but this contribution does not contain any $\log u$. This is consistent with the expectations, since the OPE contains the scalar operator of exact dimension two and all its descendants. It is straightforward to extract the anomalous dimension from \eqref{cpwa_anyd}, both for $d=2$ and $d=4$:
\es{p3gamma}{d=2:&\quad\g^{\1, \phi^3}_{n,\ell}=\left\{
                \begin{array}{ll}
                  -\frac{2(5+4n)}{(1+n)(2+n)(3+2n)}\mu_3 ^2~,\,\,\,\,\text{$\ell$=0}\\
                   -\frac{4}{(\ell+1+n)(\ell+2+n)}\mu_3^2~, \,\,\,\,\,\,\text{$\ell \neq $0}\\
                \end{array}
              \right.\\
d=4:&\quad \g^{\1, \phi^3}_{n,\ell}=\left\{
                \begin{array}{ll}
                  \frac{2(2-7(1+n)^2)}{(1+n)(3+4n(2+n))} \mu_3^2~,\,\,\,\,\text{$\ell$=0}\\
                   -\frac{8}{(\ell+1)(\ell+2+2n)} \mu_3^2~, \,\,\,\,\,\,\text{$\ell \neq $0}\\
                \end{array}
              \right.}

\bibliographystyle{utphys}
\bibliography{refs}

\end{document}